\documentclass[11pt]{article}

\usepackage[latin1]{inputenc} 
\usepackage[nosort]{cite}
\usepackage[bulletsep]{collref}
\usepackage{graphicx}
\usepackage{color} 
\usepackage{bbm}
\usepackage{bbold}
\usepackage{amsmath}
\usepackage{amssymb}
\usepackage{subfigure}
\usepackage[margin=2cm]{caption}
\usepackage{array}
\usepackage{leftidx}

\usepackage[a4paper,includeheadfoot,hmargin=2.5cm,vmargin=2cm,headheight=14pt,headsep=2.5ex,footskip=5.5ex]{geometry}

\numberwithin{equation}{section}

\makeatletter
\let\old@makecaption=\@makecaption
\def\@makecaption{\small\old@makecaption}
\makeatother

\newcommand{\unit}{\mathbbm{1}}

\newcommand{\order}{\mathcal{O}}

\newcommand{\Op}[1]{\Phi_{{#1}}}

\newcommand{\sfrac}[2]{{\textstyle\frac{#1}{#2}}}
\newcommand{\half}{\sfrac{1}{2}}

\newcommand{\ket}[1]{\mathopen{|}#1\mathclose{\rangle}}
\newcommand{\bra}[1]{\mathopen{\langle}#1\mathclose{|}}

\newcommand{\nn}{\nonumber}

\def\<{\begin{eqnarray}}
\def\>{\end{eqnarray}}

\makeatletter
\def\mr@ignsp#1 {\ifx\:#1\@empty\else #1\expandafter\mr@ignsp\fi}%
\newcommand{\multiref}[1]{\begingroup
\xdef\mr@no@sparg{\expandafter\mr@ignsp#1 \: }%
\def\mr@comma{}%
\@for\mr@refs:=\mr@no@sparg\do{\mr@comma\def\mr@comma{,}\ref{\mr@refs}}%
\endgroup}
\makeatother

\newcommand{\hypref}[2]{\ifx\href\asklfhas #2\else\href{#1}{#2}\fi}

\newcommand{\secref}[1]{Sec.~\multiref{#1}}

\newcommand{\appref}[1]{App.~\multiref{#1}}

\newcommand{\figref}[1]{Fig.~\multiref{#1}}
\renewcommand{\eqref}[1]{(\multiref{#1})}

\ifx\href\asklfhas\newcommand{\href}[2]{#2}\fi

\usepackage[bookmarks=true,hyperfigures=true]{hyperref}

\providecommand{\hypersetup}[1]{}

\providecommand{\texorpdfstring}[2]{#1}
\providecommand{\pdfbookmark}[3][]{}

\hypersetup{plainpages=false}
\hypersetup{pdfpagemode=UseOutlines}
\hypersetup{bookmarksnumbered=true}
\hypersetup{bookmarksopen=true}
\hypersetup{pdfstartview=FitH}
\hypersetup{colorlinks=false}
\hypersetup{citebordercolor={.6 .9 .6}}
\hypersetup{urlbordercolor={.7 .8 1}}
\hypersetup{linkbordercolor={1 .7 .7}}
\hypersetup{pdfborder={0 0 .5}}

\begin{document}

\thispagestyle{empty}

\begin{flushright}\footnotesize
\texttt{
TCDMATH 17-18 
}
\end{flushright}
\vspace{1cm}

\begin{center}%
{\Large\textbf{\mathversion{bold}%
Diagonal Form Factors in Landau-Lifshitz Models 
}\par}

\vspace{1.5cm}

\textrm{Lorenzo Gerotto and Tristan Mc Loughlin} \vspace{8mm} \\
\textit{%
School of Mathematics \& Hamilton Mathematics Institute \\
 Trinity College Dublin \\
Dublin, Ireland 
} \\
\texttt{gerottol@tcd.ie, tristan@maths.tcd.ie }

\par\vspace{14mm}

\textbf{Abstract} \vspace{5mm}

\begin{minipage}{14cm}
We perturbatively study form factors in the Landau-Lifshitz model and the generalisation originating in the study of the $\mathcal{N}=4$ super-Yang-Mills dilatation generator. In particular we study diagonal form factors which have previously been related to gauge theory structure constants. For the Landau-Lifshitz model, due to the non-relativistic nature of the theory, we are able to compute all orders in perturbation theory and to resum the series to find quantum form factors for low numbers of external particles. 
We apply our form factors to the study of deformations of the integrable theory by means of form factor perturbation theory. As a check of our method we compute spin-chain S-matrix elements for the  Leigh-Strassler family  of  marginal  deformations to leading order in the deformation parameters.  
\end{minipage}

\end{center}

\newpage
\tableofcontents

\vspace{10mm}
\hrule
\vspace{5mm}


\section{Introduction}
The Landau-Lifshitz (LL) model \cite{landau1935theory} was originally introduced to describe the distribution of magnetic moments in a ferromagnet and includes the Heisenberg ferromagnet equation as a special case:
\<
\frac{\partial \vec{n}}{\partial t}=\vec{n}\times \frac{\partial^2 \vec{n}}{\partial x^2}
\> 
where $\vec{n}(x,t)$ is a three-dimensional vector living on the unit sphere, $\vec{n}\cdot \vec{n}=1$.  In large part because it was found to be integrable \cite{takhtajan1977integration}, this model has subsequently been the focus of a great deal of interest in a number of different contexts. It has  played a significant role in the study of the AdS/CFT correspondence where it acted as a partial bridge between the spin-chain and string descriptions of gauge invariant operators. For small values of the 't Hooft coupling, $\lambda$, the ${\cal N}=4$ super-Yang-Mills (SYM) dilatation generator, $D$, can be computed in perturbation theory, so that 
\<
\label{eq:dilexp}
D=\sum_{r=0}^{\infty} \left(\tfrac{\lambda}{16\pi^2}\right)^r D_{2r}~.
\>
Acting on single trace operators composed of just two types of complex scalars, an $\mathfrak{su}(2)$ sub-sector of the full theory, the one-loop part can be mapped to the Heisenberg XXX$_{1/2}$ spin-chain Hamiltonian \cite{Minahan:2002ve}:
\<
\label{eq:dil2}
D_2=2 \sum_{\ell=1}^{L}(\unit -P_{\ell,\ell+1})~,
\>
where $P_{\ell,\ell+1}$ is the permutation operator acting on sites $\ell$ and $\ell+1$. 
In the thermodynamic limit the low-energy excitations about the ferromagnetic vacuum are described by an effective two-dimensional LL action \cite{Fradkin:1991nr, Kruczenski:2003gt} and the same LL action can be found as the so-called ``fast-string" limit of the bosonic string action on $\mathbbm{R}\times$S$^3$ \cite{Kruczenski:2003gt, Kruczenski:2004kw, Kazakov:2004qf}. This proved a useful tool in developing the understanding of the match between the energies of on-shell string states and anomalous dimensions at this order. Generalisations of the LL action describing larger sectors of the gauge theory were studied in \cite{Hernandez:2004uw, Stefanski:2004cw,Kruczenski:2004cn, Hernandez:2004kr, Stefanski:2005tr} and a $\mathfrak{psu}(2,2|4)$ LL model arising from the thermodynamic limit of the complete one-loop ${\cal N}=4$ SYM dilatation generator was constructed in \cite{Stefanski:2007dp}. 

Extending beyond leading order in $\lambda$, which corresponds to  considering a spin-chain Hamiltonian with longer range interactions, results in a generalised LL action with higher-derivative terms. The effective LL action to $\order(\lambda^2)$ was found in \cite{Kruczenski:2004kw} however beyond $\order(\lambda^2)$ the LL action following from the spin-chain and string theory disagree. The ``gauge"-LL action to order $\lambda^3$ was found in \cite{Minahan:2005qj} by including all six-derivative terms allowed by symmetries and fixing the coefficients by matching with the energies of known solutions and was shown to disagree with the ``string"-LL action following from the fast-string limit (see also \cite{Ryzhov:2004nz, Tseytlin:2004xa, Minahan:2005mx, Tirziu:2006ve}).
	
The LL model and its generalisations can of course be considered as two-dimensional integrable quantum field theories in their own right and their quantisation studied. The quantisation of the anisotropic LL model was studied by means of the quantum inverse scattering method and involves a number of subtleties \cite{sklyanin1988quantization}. An alternative approach is to formally introduce a small parameter and perform a perturbative calculation \cite{Minahan:2005mx, Minahan:2005qj,  Tirziu:2006ve}. This can be efficiently carried out by using the Feynman diagrammatic expansion, with the small parameter acting as a loop counting parameter, and then attempting to resum all the resulting diagrams. The quantum S-matrix for the LL model was computed in this fashion in \cite{Klose:2006dd} and generalised in  \cite{Roiban:2006yc} to include higher-order $\lambda$ corrections. In an integrable theory it is expected that the three-particle S-matrix  should factorise into the product of two-particle S-matrices, however due to the subtleties of the LL model this is non-trivial and has only been explicitly demonstrated at one-loop \cite{Melikyan:2008cy}, see also \cite{Melikyan:2008ab}. 

One advantage of the using the LL action to study the connection between the spin-chain and string theory descriptions is that it goes beyond strictly on-shell quantities such as the S-matrix. We will be interested in form factors, which are hybrid objects given by matrix elements of local operators, ${\cal O}(t,x)$,
\<
F(\theta'_m, \dots, \theta_1'|\theta_1, \dots, \theta_n)=\langle \theta'_1,  \dots, \theta_m'|{\cal O}(0,0)|\theta_1, \dots, \theta_n\rangle~,
\>
between asymptotic states. \footnote{Here the asymptotic particles are labelled by the particle rapidities $\theta_i$, $i=1,\dots, n$, and we distinguish between ``in"- and ``out"-states by the rapidity ordering:
\<
|\theta_1, \dots, \theta_n\rangle=\left\{ 
\begin{array}{lr}
	 |\theta_1, \dots, \theta_n\rangle^{({\rm in})}& \text{for } \theta_1>\dots > \theta_n \\
	 |\theta_1, \dots, \theta_n\rangle^{({\rm out})} & \text{for } \theta_1<\dots< \theta_n
	\end{array}\right.~.
\>}
Form factors are central to the bootstrap approach to quantum field theory and in integrable models can in principle be determined from a set of consistency conditions (\cite{Weisz:1977ii, Karowski:1978vz}, see \cite{Smirnov:1992vz} for a remnbcview). World-sheet form factors for the AdS/CFT correspondence were studied in \cite{Klose:2012ju, Klose:2013aza}; the LL action was used to explain how they could be matched to spin-chain matrix elements and consequently to structure constants of tree-level gauge theory three-point functions. One class of particularly interesting form factors are diagonal form factors where the two asymptotic states are taken to be identical i.e. $n=m$ and $\{\theta_1\dots \theta_n\}=\{\theta_1', \dots, \theta_m'\}$.
These are of interest in the context of the AdS/CFT duality as they are related to the structure constants of ``Heavy-Heavy-Light" three-point functions \cite{Bajnok:2014sza, Bajnok:2016xxu}. It was proposed in \cite{Bajnok:2014sza} that the dependence of structure constants on the length, $L$, of the heavy operators is given by finite volume diagonal form factors in integrable theories. This was confirmed at one-loop in \cite{Hollo:2015cda} and, based on the Hexagon approach \cite{Basso:2015zoa}, at higher loops in \cite{Jiang:2015bvm, Jiang:2016dsr}. More generally, diagonal form factors are related to the study of non-integrable deformations of integrable theories \cite{Delfino:1996xp} and can be used to determine the corrections to the vacuum energy, mass matrix and S-matrix.

With this context in mind, the goal of this paper is to perturbatively compute  form factors in the LL-model and its higher-order-in-$\lambda$ generalisations, following the perturbative methods of \cite{Klose:2006dd, Roiban:2006yc, Melikyan:2008cy}. Just as for the S-matrix, due to the theory's non-relativistic dispersion relation, the pertubative computations can be carried out to all-orders and resumed to find exact quantum form factors.  At leading order in $\lambda$ we compare these results to the low-momentum expansion of the spin-chain form factors extracted from the XXX$_{1/2}$ Heisenberg spin-chain \cite{Hollo:2015cda} and find good agreement. In principle, this comparison can be extended to the $\order(\lambda^2)$ results extracted from the Hexagon \cite{Jiang:2015bvm} and direct computation \cite{Petrovskii:2016sfs}, however in this case the world-sheet operator receives, as yet unknown, corrections and we are unable to find agreement even at  leading order in the small momentum expansion. Finally, we apply these form factors to the study of deformations of the LL model and as a simple test case consider the theory resulting from marginal deformations of the gauge theory.

\section{Landau-Lifshitz Model}

The Landau-Lifshitz model arises both in the fast-string limit of classical strings and in the low-energy limit of spin-chains. In the spin-chain description of gauge theory anomalous dimensions the one-loop dilatation operator in the $\mathfrak{su}(2)$ sector corresponds to the XXX$_{1/2}$ Heisenberg spin-chain Hamiltonian with an overall normalisation given by the 't Hooft coupling. This can be written in terms of the usual Pauli matrices,  $\sigma^i$, $i=1,2,3$, as
\<
H=\frac{\lambda}{16\pi^2}\sum_{\ell=1}^L (1-\vec{\sigma}_\ell\cdot\vec{\sigma}_{\ell+1})~.
\>
The derivation of the LL action starts with rewritting the spin-chain path integral in terms of tensor products of spin coherent states for each site which can be constructed by rotating a highest weight state oriented along the $z$-axis to define a state $\ket{\vec{n}}$ which has the property 
\<
\label{eq:CSmap}
\bra{\vec{n}}\sigma^i\ket{\vec{n}}=n^i~,~~~{\rm with}~~~\sum_{i=1}^{3}n^i=1
\>
where the $n^i$ can be thought of as parameterizing the $SU(2)/U(1)$ coset. The resulting action involves a sum over squares of differences of the vector $\vec{n}$ at neighbouring spin sites as well as a Wess-Zumino type term which is linear in the time derivative of $\vec{n}$. Introducing a coordinate $\sigma=\tfrac{2\pi \ell}{L}$,  $0<\sigma\leq 2\pi$, one keeps the low-energy modes in the continuum limit  by expanding the fields as $L\to \infty$ and keeping only the leading term in the derivative expansion. The resulting action is given by
\<
\label{eq:LLaction}
{\cal A}^{\rm LL}=\frac{L}{2\pi}\int d\tau \int_0^{2\pi} d\sigma \Big[ C_\tau(\vec{n})-\frac{\lambda}{8 L^2}(\partial_\sigma \vec{n})^2\Big]
\>
where the WZ term in this limit can be written as
\<
C_\tau = -\half \int_0^1 dz ~ \epsilon_{ijk} n^i \partial_z n^j \partial_\tau n^k~.
\>
We note that in this limit $\lambda$ only appears in the combination $\tilde \lambda=\lambda/L^2$ which is a manifestation of so-called BMN scaling \cite{Berenstein:2002jq}. With $\tilde \lambda$ held fixed, the factor of $L$ appearing in front of the action plays the role of $\hbar$ and so the tree-level results correspond to $L\to \infty$. Quantizing the theory and including loop effects corresponds to including finite $L$ corrections; however, as we have also dropped higher derivative terms in our expansion, it is not possible to recover the complete finite-$L$ result of the spin-chain via this method. 

\subsection{Perturbative Quantisation}

In order to perform a perturbative expansion following \cite{Klose:2006dd} and \cite{Roiban:2006yc, Melikyan:2008cy} we rewrite the action in two steps. Firstly, as we are interested in computing the two-dimensional S-matrix and form factors which naturally live in the two-dimensional plane rather than on the cylinder we will take a decompactification limit $L\to \infty$ while keeping $\lambda$ fixed.  Hence we rescale the spatial coordinate so that it has period $L$ and we rescale the time coordinate to simplify our expressions:
\<
\label{eq:xtcoord}
x=\frac{L\sigma}{2\pi}~, ~~~ t=\frac{\lambda \tau}{8\pi^2}~.
\>
Secondly, as was done in the Hamiltonian perturbation expansion \cite{Minahan:2005mx}, and used in computing the LL S-matrix  \cite{Klose:2006dd, Roiban:2006yc, Melikyan:2008cy}, it is convenient to solve the constraint 
$\vec{n}\cdot \vec{n}=1$ by introducing a complex field $\varphi$ given by 
\<
\label{eq:varphi}
\varphi=\frac{n^1+i n^2}{\sqrt{2+2 n^3}}~, ~~ |\varphi|^2=\frac{1}{2}(1-n^3)
\>
which is valid away from the point $n^3=-1$. An advantage of this particular transformation is that it generates an action with a canonical kinetic term:
\<
{\cal A}^{\rm LL}=\int dt \int_0^L dx \Big[\frac{i}{2}(\varphi^\ast \partial_t \varphi-\varphi \partial_t \varphi^\ast)-|\partial_x \varphi|^2-V(\varphi)\Big]\nn\\
\text{where}~~~ V(\varphi)=\frac{2-|\varphi|^2}{4(1-|\varphi|^2)}\big[(\varphi^\ast \partial_t\varphi)^2+c.c\big]
+\frac{|\varphi|^4 |\partial_x \varphi|^2}{2(1-|\varphi|^2)}~.
\>
The only dependence of the action on $L$ is now in the range of integration and we can take the decompactification limit. As the potential is clearly quite non-linear in $\varphi$, we will consider quantizing this theory near the $\varphi=0$,  i.e. $n^3=1$, vacuum by expanding the action in small $\varphi$. In the gauge theory this vacuum is given by the BPS state Tr$(Z^L)$ while in the string theory this corresponds to expanding about the BMN vacuum \cite{Berenstein:2002jq}. 

Due to the non-relativistic form of the quadratic action the field $\varphi(t,x)$ can be expanded in negative energy modes only\footnote{Our normalisation of the creation and annihilation operators is the same as \cite{Klose:2006dd, Melikyan:2008cy} and differs from \cite{Roiban:2006yc} by $\sqrt{2\pi}$.} 
\<
\varphi(t,x)=\int \frac{dp}{2\pi}~ a_p e^{-i\omega_p t+ipx}
\>
where the particle energy $\omega_p=p^2$ and the conjugate field is given by 
\<
\varphi^\ast(t,x)=\int \frac{dp}{2\pi}~ a^\dagger_p e^{i\omega_p t-ipx}~.
\>
The operators $a_p$ and $a^\dagger_p$ are annihilation and creation operators for particles of momentum $p$ and satisfy the usual commutation relations 
\<
[a_p,a^\dagger_{p'}]=2\pi \delta(p-p')~,
\>
and the ground state is annihilated by the field operator $\varphi(t,x)\ket{0}=0$. 
An essential feature of this model, emphasised in \cite{Klose:2006dd}, is that due to the non-relativistic form of the kinetic term, the propagator has a single pole in momentum space
\<
 \tilde D(\omega, p)=~~
\includegraphics[scale=1]{./LLfigures/FigProp}~~=\frac{i}{\omega-p^2+i0}
\>
and correspondingly in position space is purely retarded
\<
D(t,x)=\Theta(t)\sqrt{\frac{\pi}{it}}~\text{exp}\left(\frac{i x^2}{4 t}\right)~.
\>
This results in a number of important simplifications in the perturbative calculation, in particular the direction of the arrow on the propagator is essential as any diagram with a closed loop containing propagators whose arrows point in the same direction vanishes. This implies the non-renormalisation of the vacuum energy and one-particle propagator. Furthermore, the two-body S-matrix is given by a sum of bubble diagrams; as we will see, a similar simplification occurs for form factors.    
\subsection{Generalised Landau-Lifshitz Model}
We will study the generalisation of this model to include the higher-order-in-$\lambda$ corrections. The two-loop, $\order(\lambda^2)$, LL model was studied in \cite{Kruczenski:2004kw}, the three-loop, $\order(\lambda^3)$, in \cite{Minahan:2005mx, Minahan:2005qj} and the four-loop, $\order(\lambda^4)$, in \cite{Tirziu:2006ve}. In all cases the expansion organises itself such that the action can be written in terms of $\tilde \lambda$ and the remaining dependence on the spin-chain length is an overall factor. We will restrict ourselves to the three-loop expressions:  
\<
{\cal A}^{\rm gLL}&=&\frac{L}{2\pi}\int d\tau \int_0^{2\pi} d\sigma \Big[ C_\tau(\vec{n})-\frac{ \tilde \lambda b_0}{8 }(\partial_\sigma \vec{n})^2
-\frac{\tilde \lambda^2}{32}(b_1 (\partial^2_\sigma \vec{n})^2+b_2 (\partial_\sigma \vec{n})^4)\nn\\
& &
\kern+20pt -\frac{\tilde \lambda^3}{64}\left(b_3 (\partial^3_\sigma \vec{n})^2+b_4 (\partial_\sigma \vec{n})^2 (\partial^2_\sigma \vec{n})^2
+b_5 (\partial_\sigma \vec{n}\cdot \partial^2_\sigma \vec{n} )^2 +b_6 (\partial_\sigma \vec{n})^6\right)
\Big]~.
\>
Here we have left the coefficients $b_1, \dots, b_6$ arbitrary however they can be fixed by computing the energies of specific solutions and comparing with known gauge theory and string theory results. The string theory and gauge theory values are the same for the coefficients up to $\order(\lambda^2)$ 
\<
b_0=1,~b_1=-1,~  \text{and } b_2=\tfrac{3}{4}~.
\>
At the next order the value of $b_3$ is fixed to be $1$ by demanding BMN-like scaling for the magnon energy while in order to reproduce the known gauge theory anomalous dimensions to $\order(\lambda^3)$ the required values are 
\<
b_4=-\tfrac{7}{4},~  
b_5=-\tfrac{23}{2},~ \text{and } b_6=\tfrac{3}{4}~.
\>
To match with the string results we have the same value for  $b_4$ but
\<
b_5=-\tfrac{25}{2},~ \text{and } b_6=\tfrac{13}{16}~.
\> 
Importantly for our purposes the kinetic term $C_\tau(\vec{n})$ does not receive any corrections. 
Rescaling the world-sheet coordinates as above, \eqref{eq:xtcoord}, introducing the complex scalar field $\varphi$, defining the parameter $g=\tfrac{\sqrt{\lambda}}{4\pi}$ and expanding in powers of the field we find the action
\<
\label{eq:genLLaction}
{\cal A}^{\rm gLL} = \int d^2x& &  \kern-15pt \Big\{ \frac{i}{2}(\varphi^\ast \partial_t \varphi-\varphi \partial_t \varphi^\ast)-b_0|\partial_x \varphi|^2-g^2 b_1   |\partial_x^2 \varphi|^2  -2 g^4 b_3   |\partial_x^3 \varphi|^2  
\nn\\
& &-  V_{\rm quartic}-V_{\rm sextic} +\cdots \Big\}
\>
where the terms in the potential are to leading order
\<
V_{\rm quartic}&=&   \frac{b_0}{2}(\varphi^\ast{}^2 (\partial_x \varphi)^2+\varphi^2(\partial_x \varphi^\ast)^2)+\order(\lambda)\nn\\
V_{\rm sextic}&=&-\frac{b_0}{4}\varphi \varphi^\ast(\varphi^\ast \partial_x \varphi +\varphi \partial_x \varphi^\ast)^2+\order(\lambda)
\>
and we give the higher order terms in \appref{app:HOPot}. The form of the potential is not identical to that of \cite{Roiban:2006yc} however the difference is due to total derivative terms and, as we will check below, gives rise to the same S-matrix. As at leading order, the spin-chain length now only appears in the range of integration and so we can again take the decompactification limit. However, the rescaling of the time coordinate does not remove the dependence on $\lambda$ (now $g$) which now appears explicitly even at quadratic order. This results in a modification of the dispersion relation in addition to new, higher derivative, interaction terms. 

\subsection{Feynman Rules}
While the quadratic higher-order-in-$g$ terms result in a corrected dispersion relation 
\<
\omega(p)=b_0 p^2+g^2 b_1 p^4+2 g^4 b_3 p^6
\>
the corresponding propagator still only has a single pole 
\<
\label{eq:prop}
\tilde D(\omega, p)=\frac{i}{\omega-b_0 p^2-g^2 b_1 p^4-g^4 b_3 p^6+i0}
\>
and so remains purely retarded. This ensures that we have the same non-renormalisation theorems and simplifications in the diagrammatic expansion as in the leading-order LL model. For example the quantum S-matrix is still simply given by a sum over bubble diagrams, \cite{Roiban:2006yc}, but now with more complicated  vertices. 

The quartic vertex is 
\<
\label{eq:quart_vert}
\vcenter{ \includegraphics[scale=0.75]{./LLfigures/FigQuartic.mps}}\kern-380pt &: &2i b_0 (k_1 k_2+p_1 p_2)-2 ig^2 \Big[4(3b_1+2 b_2) p_1 p_2 k_1 k_2\\
& & +b_1 \left(k_1^2 k_2^2+p_1^2 p_2^2 -2(p_1+p_2)(k_1+k_2)(k_1 k_2+p_1 p_2)\right)\Big]\nn\\
& & \kern-0pt
~-2ig^4
\Big[(2 b_4 +b_5) ((k_1+k_2)(p_1+p_2)-2 b_5 (k_1 k_2+p_1 p_2))k_1 k_2 p_1 p_2\nn\\
& &\kern+30pt
+b_3\left(3( k_1 k_2 p_1 p_2(2 (k_1^2+k_2^2+p_1^2+p_2^2)+6(k_1+k_2)(p_1+p_2)\right.
\nn\\
& &\kern+30pt  -12 (k_1 k_2+p_1 p_2))+
(k_1^2 k_2^2+p_1^2 p_2^2)(p_1+p_2)(k_1+k_2)
\nn\\
& &\kern+30pt -(k_1 k_2+p_1 p_2)(p_1^2+p_2^2)(k_1^2+k_2^2)
-(k_2^3+k_1^3)(p_1+p_2)p_1 p_2\nn\\
& &\kern+30pt \left. -(p_2^3+p_2^3)(k_1+k_2)k_1 k_2\right)-2 k_1^3 k_2^3-2 p_1^3 p_2^3)\Big]\nn
\>
where there is understood to be an overall momentum conservation delta-function imposing $p_1+p_2=k_1+k_2$. 
Finally we will also make use of the sextic vertex to calculate the three-particle S-matrix and form factors:
\<
\vcenter{		\includegraphics[height=0.075\textheight]{./LLfigures/FigSextic.mps}}\kern-380pt & &:
 i b_0  \Big[3(k_1 k_2 +k_1 k_3+k_2 k_3)+3 (p_1 p_2 +p_1 p_3+p_2 p_3)
 \nn\\
 & &\kern+40pt -2(k_1+k_2+k_3)(p_1+p_2+p_3)\Big]+\order(g^2)
\>
where we have only written the leading term in $\order(g^2)$. The subleading terms can be extracted straighforwardly from the sextic potential \eqref{eq:sextic_pot}.

\subsection{Generalised Landau-Lifshitz S-matrix}

The calculation of the quantum S-matrix from the quartic vertex was carried out for the leading-order LL model in \cite{Klose:2006dd} and was done for the generalised LL-model in \cite{Roiban:2006yc}. Here we briefly recap this calculation as it both provides a check on the form of our action and is closely related to that of form factors. The two-body S-matrix is defined by 
\<
\label{eq:Smat_def}
\bra{k_1 k_2}{\hat S}\ket{p_1 p_1}=\bra{k_1 k_2}{\rm T}\big[\text{exp}\left(-i\int d^2 x~ H_I \right)\big]\ket{p_1 p_2}
\>
where $H_I$ is the interaction Hamiltonian, the asymptotic states are given simply by 
\<
\ket{p_1 p_2}=a_{p_1}^\dagger a_{p_2}^\dagger\ket{0}~, ~~~ \bra{k_1 k_2}=\bra{0}a_{k_1}a_{k_2}
\>
and in the perturbative expansion we only keep amputated, connected terms. Due to spatial- and time-translational invariance of the action the S-matrix elements, \eqref{eq:Smat_def}, naturally come with overall energy, $p^0_i=\omega(p_i)$, and momentum, $p^1_i=p_i$, delta-functions, 
\<
(2\pi)^2\delta^{(2)}(p^\mu_1+p^\mu_2-k^\mu_1-k^\mu_2)={\cal J}\delta_+(p_1,p_2,k_1,k_2)
\>
where the Jacobian factor is ${\cal J}=1/(\partial \omega(p_1)/\partial p_1-\partial \omega(p_2)/\partial p_2)$
and 
\<
\delta_+(p_1,p_2,k_1,k_2)
&=&(2\pi)^2 \left(\delta(p_1-k_1)\delta (p_2-k_2)+\delta(p_1-k_2)\delta(p_2-k_1)\right)~.
\>
We define the T-matrix, $S(p_1,p_2)=1+ T(p_1,p_2)$, where 
\<
\bra{k_1 k_2}{\hat S}\ket{p_1 p_1} = S(p_1,p_2)\delta_+(p_1,p_2,k_1,k_2)~,
\>
to include the Jacobian factor.

The action \eqref{eq:genLLaction} has an implicit small parameter, from the expansion in powers of the fields, with which we can organise a diagrammatic expansion. For the two-body S-matrix the leading term is the tree-level quartic contribution,  which gives,
\<
\label{eq:T0gLL}
T^{(0)}(p_1,p_2)&=&\frac{2 i p_1 p_2}{p_1-p_2}-\frac{2 i g^2}{b_0 }\frac{(5 b_1+4 b_2)p_1^2p_2^2}{p_1- p_2}
\\
& & +\frac{2 i g^4 p^2_1 p^2_2}{b_0^2(p_1-p_2)}\big[ 10 b_1^2 (p_1^2+p_1 p_2+p_2^2)+8 b_1 b_2 (p_1^2+p_1 p_2+p_2^2)-b_0 (b_5 (p_1-p_2)^2\nn\\
& &\kern+70pt +2 b_4(p_1+p_2)^2+7 b_3(3 p_1^2-2 p_1 p_2+3 p_2^2))\big]\nn~.
\>
Using the string values for the coefficients we find the appropriate LL limit of the known string and spin-chain S-matrices to this order in $g$ \cite{Roiban:2006yc}.

\begin{figure}\centering
	\includegraphics[height=0.06\textheight]{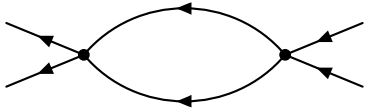}\qquad
	\caption{One-loop bubble diagram contribution to two-body S-matrix.}
	\label{fig:2b1loop}
\end{figure}
\begin{figure}
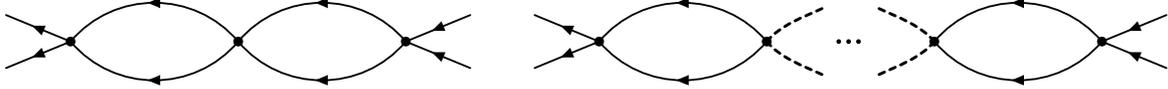
\centering
	\includegraphics[height=0.05\textheight]{./LLfigures/FigSS2Loop.mps}\qquad
	\includegraphics[height=0.05\textheight]{./LLfigures/FigSSAllLoop.mps}
	\caption{Higher loop bubble diagrams for the two-body S-matrix.}
	\label{fig:2b2loop}
\end{figure}

As was shown in \cite{Klose:2006dd}, due to the nature of the LL propagator only bubble diagrams contribute the S-matrix calculation, and these can be calculated by simple contour integration. The same arguments can be applied in 
this generalised model \cite{Roiban:2006yc}. Going to higher-orders in $g$ results in higher powers of momenta in both the propagators and numerators. However there are no additional powers of the energy, $\omega$, and so the contour argument goes through.  
We use the full vertex \eqref{eq:quart_vert} and propagator \eqref{eq:prop} before expanding
in $g$ to evaluate the diagram in \figref{fig:2b1loop}. The resulting loop integral is naively UV divergent with power-like divergences: these can be treated by use of dimensional regularisation, which for practical purposes
amounts to essentially ignoring them \cite{Klose:2006dd}. To order $g^4$, and using values for $b_i$'s that reproduce the tree-level BDS 
S-matrix we find, as in \cite{Roiban:2006yc},
\<
\label{eq:T1spin}
T_{\rm gauge}^{(1)}=-\frac{2 p^2_1 p^2_2}{(p_1-p_2)^2}(1+4  g^2 p_1 p_2 -4 g^4 p_1 p_2(p_1-p_2)^2)
\>
which as will been seen later agrees with the one-loop BDS result to ${\cal O}(g^4)$. The corresponding result 
with the string theory coefficients is quite similar but differs at ${\cal O}(g^4)$,
\<
\label{eq:T1string}
T_{\rm string}^{(1)}=-\frac{2 p^2_1 p^2_2}{(p_1-p_2)^2}(1+4  g^2 p_1 p_2 -2 g^4 p_1 p_2(p_1-p_2)^2)~.
\>
As in the leading-order calculation this can be extended to two- and higher-loop by evaluating higher loop bubble diagrams  \figref{fig:2b2loop}. Each bubble can be essentially evaluated independently and so the result is a geometric series which can be easily resummed. 

\begin{figure}
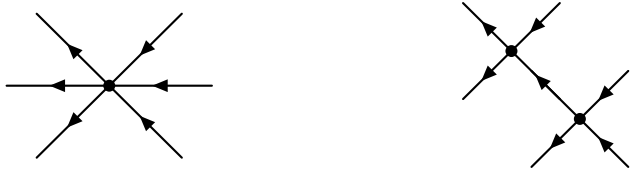
\centering 
	$\vcenter{\hbox{	\includegraphics[scale=0.9]{./LLfigures/FigSexticnl.mps}}}$\hspace{3cm}
	$\vcenter{\hbox{	\includegraphics[scale=0.9]{./LLfigures/Fig3part.mps}}}$
	\caption{Contact and dog-diagram contribution to three-body S-matrix.}
	\label{fig:3bfact}
\end{figure}

\paragraph{S-matrix Factorisation} As the theory is known to be integrable we of course expect the generalised LL model to exhibit factorised scattering. This implies that the three-body S-matrix is only non-vanishing when the out-going momenta are a permutation of the incoming momenta. For the LL model and its generalisation, as there are sextic terms in the potential, such a factorisation is not immediately apparent and results from a non-trivial cancellation between diagrams. Factorisation of scattering at one-loop for the standard LL-model was shown in \cite{Melikyan:2008cy} and was further studied in \cite{Melikyan:2008ab}. 
To check tree-level factorisation for the generalised LL-model to $\order(g^4)$ we computed the $3\to 3$ scattering by evaluating the diagrams \figref{fig:3bfact} and then checked numerically that for generic external momenta the scattering vanished to order $g^4$.
 It is interesting to note that to order $g^2$ there are no constraints on $b_1$ and $b_2$. This perhaps should be expected as the naive continuum limit of the two-loop dilatation generator and the string model differ in the value of $b_2$ and yet should both be integrable. At the next order, i.e. $g^4$, the vanishing of the 
generic $3\to 3$ S-matrix requires cancellation between terms involving different $b_i$'s. For example
fixing $b_0$ through $b_4$ as above, we find the condition $1-2b_5-32 b_6=0$. This is naturally satisfied in both the BDS case $b_5=-\tfrac{23}{2}$
and $b_6=\frac{12}{16}$ and in the string case with $b_5=-\tfrac{25}{2}$
and $b_6=\frac{13}{16}$. When the set of outgoing momenta is a permutation of the incoming momenta specific internal propagators in diagrams of the form \figref{fig:3bfact} will go on-shell and so there are additional non-vanishing contributions from delta-functions arising from using the principle value prescription
\<
\frac{1}{z+i0}=-i\delta(z)+{\rm P.V.}\Big[\frac{1}{z}\Big]~.
\> 
We now turn to the the analogous computations for form factors.

\section{Landau-Lifshitz  Form Factors} 
\label{sec:LLff}
We will focus on the computation of diagonal form factors both because of their general interest and because of their role in the AdS/CFT correspondence. Diagonal form factors can be defined, as for example in \cite{Pozsgay:2007gx, Hollo:2015cda}, as matrix
elements of operators between multi-particle states labeled by the particle rapidities $u_i$ or 
momenta $p_i$
\<
{}^{\rm in}\langle u_1,\dots , u_n|{\cal O}(0,0)|u_1, \dots, u_n\rangle^{\rm in}~.
\>
Such objects are, however, singular and require regularisation which is provided by shifting each of the rapidities in the bra-state by a small amount $u_i \to u_i+\epsilon_i$. 
In a theory with a crossing symmetry which relates a outgoing particle with rapidity $u$ to an incoming
anti-particle with rapidity $\bar u$,  such diagonal elements can be related to the usual form factors\footnote{In a relativistic theory $\bar u=u+i \pi$ however the theory need not necessarily be relativistically invariant and the shift will depend on the theory; of particular interest is the AdS/CFT world-sheet theory for which this is the case.}
\<
f^{\cal O}(\bar u_1+\epsilon_1, \dots, \bar u_n+\epsilon_n, u_1, \dots, u_n)=\langle 0|{\cal O}|\bar u_1+\epsilon_1, \dots, \bar u_n+\epsilon_n, u_1, \dots, u_n\rangle^{\rm in}~.
\>
In general the limit $\epsilon_i\to 0$ is not well defined and it was noted in \cite{Delfino:1996xp}
that the result depends on how the limit is taken, that is on the so-called scheme. In the notation of \cite{Pozsgay:2007gx} and  \cite{Hollo:2015cda}, the general 
result can be written as 
\<
\label{eq:reg_dff}
f^{\cal O}(\bar u_1+\epsilon_1, \dots, \bar u_n+\epsilon_n, u_1, \dots, u_n)&=&\prod^n_{i=1}\frac{1}{\epsilon_i}
\sum_{i_1}^n\dots \sum_{i_n}^n a_{i_1\dots i_n}(u_1, \dots, u_n)\epsilon_{i_1}\dots \epsilon_{i_n}
\nn\\
& &\kern+20pt +{\rm~ terms~ vanishing~ as}~ \epsilon_i \to 0~,
\>
where $a_{i_1\dots i_n}$ is a completely symmetric tensor. 
The symmetric scheme defines the diagonal form factor by taking all the $\epsilon_i$'s to 
be the same, $\epsilon_i=\epsilon$ for each $i=1, \dots, n$ and then setting $\epsilon\to 0$. 
Alternatively the connected scheme defines the diagonal form factors as the finite part of 
\eqref{eq:reg_dff},
\<
f^{{\cal O}}_c( u_1, \dots, u_N)\equiv n! ~a_{12\dots n}~.
\>
The LL model doesn't have crossing and thus we will instead directly calculate 
\<
F^{{\cal O}}( p_1, \dots, p_n)={}^{\rm out}\langle p_1+\epsilon_1,\dots, p_n+\epsilon_n|{\cal O}(0,0)|p_1, \dots, p_n\rangle^{\rm in}~,
\>
where we use one ``in"-state and one ``out"-state and we label our states by momenta rather than rapidities. 
In what follows we will consider taking the diagonal limit with both the symmetric and connected  prescriptions. In this limit only the zero-momentum component of our insertion operators will contribute.
From the Fourier transform
\<
 {\cal O}(t,x)=\int \frac{d^2q}{(2\pi)^2}~  \tilde{{\cal O}}(\omega_q,q) e^{iq x-i\omega_q t}~,
\>
and using the convention that the operator momentum is incoming, overall energy and momentum conservation implies 
\<
\omega_q=\sum_{i=1}^N\omega(p_i+\epsilon_i)-\omega(p_i)~~~ {\rm and}~~~ q=\sum_{i=1}^N \epsilon_i~.
\>
As both $\omega_q\to 0$ and $q \to 0$ in the diagonal limit only the zero-component will have a non-vanishing matrix element. Hence we will calculate 
\<
F^{{\cal O}}( p_1, \dots, p_n)={}^{\rm out}\langle p_1+\epsilon_1,\dots, p_n+\epsilon_n|\tilde {\cal O}(\omega_q ,q)|p_1, \dots, p_n\rangle^{\rm in}~,
\>
and then take the diagonal limit.
Additionally, because of our use of the ``out"-states in the defintion of $F^{{\cal O}}$ and the usual relation 
\<
{}^{\rm out}\langle k_1,\dots, k_M|={}^{\rm in}\langle p_1,\dots, p_N|S(p_1,\dots, p_N;k_1,\dots,k_M)
\>
we will find additional factors of the S-matrix when compared to $f^{{\cal O}}$ as computed in \cite{Hollo:2015cda}.

 \subsection{\texorpdfstring{$|\varphi|^2$}{Phi2}-Operator}
 \label{sec:length1op}
 We will start by taking the composite operator to be 
 \<
 \Op{1}=|\varphi|^2~.
 \>
 As we show below in \secref{sec:XXXFF}, this will correspond to a length-one operator in the spin-chain language. It is apparent that the vacuum expectation, or zero-particle form factor is vanishing, $F^{\Op{1}}(\emptyset)=0$, and the one-particle form factors are essentially trivial, they receive no loop corrections, and with our normalisations are given by
 \<
F^{ \Op{1}}(p)=1
\>
which corresponds to a definition of the external states without normalisation factors involving the particle energy. Both of these facts follow from vanishing of loop diagrams with arrows forming a closed loop and are correspondingly true in both the LL model and the generalisation to higher order in $g$. 

\begin{figure}\centering
	\includegraphics[height=0.125\textheight]{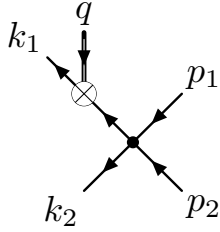}\qquad
	\caption{Tree-level contribution to length-one two-particle form factor.}
	\label{fig:L1FF0}
\end{figure}

For the two-particle form factors $F^{|\varphi|^2}(p_1,p_2)$, however, we have non-trivial results. 
Starting with the tree level, we must evaluate the diagram in \figref{fig:L1FF0}
which at $\order(g^0)$ gives
\<
 \frac{-2 b_0 \big[(k_1-q)k_2+p_1 p_2 \big]}{\omega_{k_1}-\omega_q-b_0 (k_1-q)^2 + i0}
\>
which is clearly singular in the diagonal limit. However after summing over the diagrams with the insertion on the other legs the limit becomes regular and one finds
\<
\left. F_{s,c}^{(0)  \Op{1}}(p_1,p_2)\right|_{\order(g^0)}&=&-2\kappa_{s,c}\frac{p_1^2+p_2^2}{(p_1-p_2)^2}~.
\>
This result is the same regardless of whether it is calculated using the symmetric or the 
connected prescription up to the overall normalisations. In the connected prescription one finds
$\kappa_c=1$ while in the symmetric prescription it is $\kappa_s=2$. 
Given the corrected propagator and vertex for the generalised LL-model we can extend this to 
higher orders in $g^2$: 
\<
F_{s,c}^{(0)  \Op{1}}(p_1,p_2)&=&2\kappa_{s,c} \Big[-\frac{p_1^2+p_2^2}{(p_1-p_2)^2}
+ g^2 \frac{2(4 b_2 +5 b_1) p_1 p_2(p_1^2-p_1 p_2+p_2^2)}{b_0 (p_1-p_2)^2}\nn\\
& & \kern-80pt
+ g^4 \frac{2 p_1 p_2}{b_0^2 (p_1-p_2)^2}
\big[(-10b_1-8 b_1 b_2+b_0 (21 b_3+2 b_4+ b_5)) (p_1^4+p_2^4)\nn\\
& &
+(5b_1^2+4 b_1 b_2+b_0(-63 b_3+2 b_4-5 b_5))( p^3_1 p_2+p_1 p^3_2)\nn\\
& &+(-20 b_1^2-16 b_1 b_2 +8 b_0(14 b_3 +b_5))p_1^2 p_2^2
\big]\Big]
\>
where the values of the coefficients for the different prescriptions, $\kappa_{s,c}$, are as above. 
Using the specific choices for the coefficients $b_i$ we find  that in the connected prescription
\<
\label{eq:ff2c}
F_c^{(0)  \Op{1}}(p_1,p_2)&=&-\frac{2(p_1^2+p_2^2)}{(p_1-p_2)^2}
-\frac{8 g^2 p_1 p_2 (p_1^2-p_1 p_2+p_2^2)}{(p_1-p_2)^2}
\>
to order ${\cal O}(g^2)$ for both the string theory and BDS gauge theory cases while 
\<
\left. F_c^{(0) \Op{1}}(p_1,p_2)\right|_{{\cal O}(g^4)}&=& \frac{4 p_1 p_2}{(p_1-p_2)^2}
\times \begin{cases}
 (p_1^4-2p_1^3 p_2+4p_1^2p_2^2-2p_1 p_2^2+p_2^4), ~~~{\rm string~ case}
\\
 (2p_1^4-7p_1^3 p_2+12p_1^2p_2^2-7p_1 p_2^2+2p_2^4)~, ~~~{\rm gauge~ case}
\end{cases}\nn
\>
reflecting the three-loop difference at the level of the form factor. 

\begin{figure}
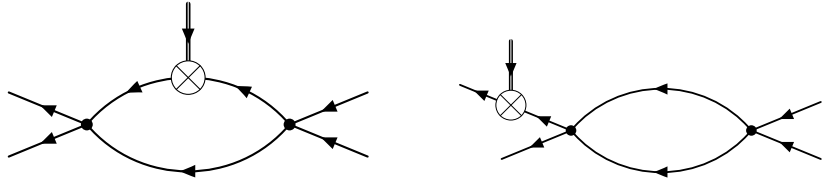
\centering
\includegraphics[width=0.3\textwidth]{./LLfigures/FigFF1Loop_int.mps}\quad \qquad
\includegraphics[width=0.3\textwidth]{./LLfigures/FigFF1Loop_ext.mps}
\caption{One-loop diagrams for two-particle, length-one form factor.}
\label{fig:1loop2part}
\end{figure}

\paragraph{One-loop result}
In order to compute the one-loop results we must consider the diagrams shown in \figref{fig:1loop2part}.
The procedure for evaluating these diagrams is essentially identical to that used in the case of the S-matrix. We regularize any power-like divergences by dimensional regularization and evaluate the integrals by using the residue theorem. In simplifying our formulas we explicitly assume that $p_1>p_2$. The same assumption will be made in all loop-calculations that we perform for form factors. 
The choice of the prescription for taking the diagonal limit superficially appears to make a more significant difference at loop level as there are different contributions from individual diagrams.
 Using the symmetric prescription we find that diagrams with the form factor inserted 
 on external legs and internal legs contribute equally so that for the LL-model i.e. to ${{\cal O}(g^0)}$
 we find
 \<
\left. F^{(1) \Op{1}}_{s,c}(p_1,p_2)\right|_{{\cal O}(g^0)}&=&- \frac{4\kappa_{s,c}  i p_1 p_2 (p_1^2+p_2^2)}{(p_1-p_2)^3}~.
\>
In the connected prescription the diagrams with the insertion on the external 
legs do not contribute at all but the diagrams with the insertion on the internal legs contributes the same as in the symmetric case. Hence we find that the connected scheme gives half of symmetric result which just as at tree-level.

 \paragraph{All-loop result}
 
To extend these results to all-loop we need only to consider chains of bubble diagrams. There are again essentially two classes of diagrams: those  with the insertion on the external leg and those with the insertion on an internal loop leg. For each bubble we can perform the loop integration by evaluating the residues. For the symmetric prescription we find equal contributions from the insertions on the external legs and from the $n$ internal insertions with the final 
 result that at $n$-loops we have,
  \<
\left. F^{(n)  \Op{1}}_s(p_1,p_2)\right|_{{\cal O}(g^0)} &=& 
4(n+1)\frac{   i^{n+2} p^n_1 p^n_2 (p_1^2+p_2^2) }{ (p_1-p_2)^{n+2} }~.
\>
Alternatively, for the connected prescription we find for the contribution with the insertion on the external legs
\<
2(n-1) \frac{  i^{n} p ^n_1 p^n_2 (p_1^2+p_2^2) }{ (p_1-p_2)^{n+2} }~.
\>
Taking the connected diagonal limit for the diagrams with internal insertions is slightly complicated
but it can be numerically checked that it gives
\<
4n\frac{    i^{n+2} p^n_1 p_2^n (p_1^2+p_2^2) }{ (p_1-p_2)^{n+2} }~
\>
so that
  \<
\left. F^{(n)  \Op{1}}_c(p_1,p_2)\right|_{{\cal O}(g^0)} &=& 
2(n+1)\frac{   i^{n+2} p^n_1 p^n_2 (p_1^2+p_2^2) }{ (p_1-p_2)^{n+2} }~,
\>
which is again simply half the symmetric prescription.
In both cases we can sum up the contributions from each loop order to give the all-loop quantum form factor:
\<
\label{eq:LL2parto1}
\left. F^{ \Op{1}}_{s,c}( p_1,p_2)\right|_{{\cal O}(g^0)} &=& -
\frac{2\kappa_{s,c} (p_1^2+p_2^2) }{ (p_1-p_2)^{2} }\frac{1}{\left(1-\frac{ip_1p_2}{p_1-p_2}\right)^2}~.
\>

 \paragraph{Three-particle form factor}
  \begin{figure}
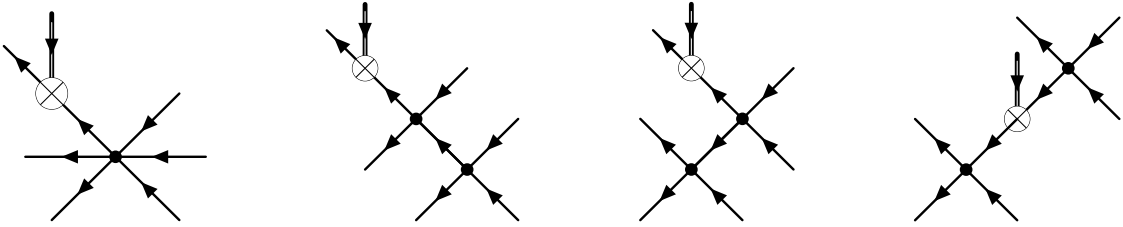
\centering
  	
  	\includegraphics[scale=0.95]{./LLfigures/FigFFsextic.mps}\qquad\qquad
 	\includegraphics[scale=0.95]{./LLfigures/FigFF3ptdoga.mps}\qquad \qquad
 	\includegraphics[scale=0.95]{./LLfigures/FigFF3ptdogb.mps}\qquad\qquad
 		\includegraphics[scale=0.95]{./LLfigures/FigFF3ptdogint.mps}
 	\caption{Diagrams for three-particle, length-one form factor.}
 	\label{fig:3partL1}
 \end{figure}
It is straightforward, if somewhat cumbersome, to extend to higher numbers of particles in the external states. In this case we must include the contributions from the graphs in \figref{fig:3partL1}. From a perturbative perspective this is of interest as it includes contributions from the sextic vertex. Furthermore in this case the dependence on the prescription for taking the diagonal limit is more pronounced. It is convenient to define the functions of external momenta
\<
p_{ij}=p_i-p_j~,~~~\chi_{i,j,k}=p_i p_j-p_j p_k+p_k p_i
\>
such that for the connected prescription the result can be written as 
\<
\left. F^{ \Op{1}}_c(p_1,p_2,p_3)\right|_{{\cal O}(g^0)} &=&
 \frac{4}{p_{12}^2 p_{13}^2 p_{23}^2}\Big[p_1^4(p_2^2+p_3^2)+p_2^4(p_1^2+p_3^2)+p_3^4(p_1^2+p_2^2)\nn\\
 & &-2 p_1p_2p_3(p_1 \chi_{1,2,3}+ p_2 \chi_{2,3,1}+ p_3 \chi_{3,1,2})\Big]
 \>
 while the symmetric prescription gives
 \<
 \left. F^{ \Op{1}}_s(p_1,p_2,p_3)\right|_{{\cal O}(g^0)} &=&
 \frac{24}{p_{12}^2 p_{13}^2 p_{23}^2}\Big[(p_1^2 p_2^2+p_3^2p_1^2+p_2^2p_3^2)(p_1^2+p_2^2+p_3^2-p_1 p_2-p_3p_1-p_2p_3)\Big]~.\nn\\
 \>
  
\paragraph{Further Quadratic Operators}
 It is possible to consider other quadratic-in-field operators by adding derivatives. 
One such operator which will be relevant to our later considerations is 
 \<
\Op{2}= \varphi^\ast \acute{\varphi}-\varphi \acute{\varphi}^\ast
 \>
 for which one can calculate the results $ F^{\Op{2}}_{c,s}(\emptyset)=0$, $ F^{\Op{2}}_{c,s}(p_1)=2ip_1$ 
 and
  \<
 \left. F^{\Op{2}}_{c,s}(p_1,p_2)\right|_{{\cal O}(g^0)} &=&
 -\frac{4i\kappa_{c,s} p_1 p_2}{(p_1-p_2)^2}\frac{(p_1+p_2)}{\left(1-\frac{i p_1p_2}{p_1-p_2}\right)^2}~.
 \>
 One feature of this calculation is that as the insertion operator involves derivatives when it is inserted inside a loop, as in \figref{fig:1loop2part}, it gives rise to additional numerator factors. In this case care must be taken in the labelling of the loop momenta passing through the insertion. In particular we must sum over contributions corresponding to inserting the operator on the top line with loop momentum $\ell$ and the bottom line with momentum $-\ell+p_1+p_2$ as these are not equal. 
 
There are of course many other possible operators one could consider. If there were two derivatives such terms could act as possible higher order corrections to the $|\varphi|^2$ operator, for example
  \<
  {\cal O}_{\rm corr}=|\varphi|^2+g^2 \Big[\alpha_1 (\partial_x^2\bar \varphi^\ast)\varphi+
  \alpha_2 \bar \varphi^\ast(\partial_x^2 \varphi)+\alpha_3 (\partial_x \bar \varphi^\ast)(\partial_x \varphi)\Big]~.
  \>
Of course as the correction terms are related by total derivatives, for diagonal form factors we would
expect the three correction terms to give the same contributions and so there is only one parameter at this order. As we will see, such corrections are likely to play a role in understanding the relation to gauge theory structure constants.  

  \subsection{\texorpdfstring{$|\varphi|^4$}{Phi4}-Operator}
  \begin{figure}\centering
  	\includegraphics[width=0.11\textwidth]{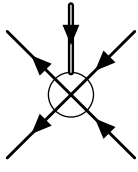}
  	\caption{Length-two two-particle tree-level form factor diagram.}
  	\label{fig:FFl22part}
  \end{figure}
We now turn to the  $|\varphi|^4$ operator which will correspond to a length two-operator in the spin-chain language. The zero-particle form factor is again obviously vanishing as is the one-particle diagonal form factor. The two particle diagonal form factor at tree-level is simply 
 \<
\left. F^{(0) \Op{3}}(p_1,p_2)\right|_{{\cal O}(g^0)}&=& 4
 \>
corresponding to \figref{fig:FFl22part}. The loop corrections in the LL-model are given by essentially the same diagrams as in the S-matrix calculation, \figref{fig:2b2loop}, with one of the interaction vertices replaced by the operator insertion. These diagrams can again be resummed to give 
 \<
\left. F^{\Op{3}}(p_1,p_2)\right|_{{\cal O}(g^0)}&=& \frac{4}{\left(1-\frac{ip_1 p_2}{p_1-p_2}\right)^2}~
\> 
 where the result does not depend on the prescription used in taking the diagonal limit. 
 \begin{figure}
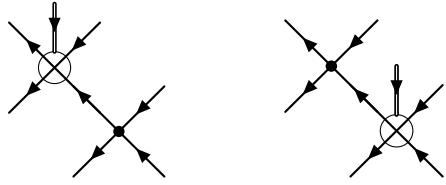
\centering
 	\includegraphics[height=0.1\textheight]{./LLfigures/Fig3partalength2.mps}\qquad\qquad
 	\includegraphics[height=0.09\textheight]{./LLfigures/Fig3partblength2.mps}
 	\caption{Length-two three-particle tree-level form factor diagrams.}
 	\label{fig:FFl23part}
 \end{figure}

 \paragraph{Three-particle form factor}
 For the  $|\varphi|^4$ operator it is particularly straightforward to extend to three-particles by evaluating the diagrams shown in \figref{fig:FFl23part}. However now the result does depend on the presciption used to define the diagonal limit in much the same fashion as the two-particle form factors of  $|\varphi|^2$.
  \<
\left. F^{0,\Op{3}}_{s,c}(p_1,p_2,p_3)\right|_{{\cal O}(g^0)}&=&8(\kappa_{s,c}-1) -\frac{8 \kappa_{s,c}}{p_{12}^2p_{23}^2 p_{31}^2} \big[p_{31}^4\chi_{2,3,1}
 +p_{23}^4 \chi_{1,2,3}+p_{12}^4\chi_{3,1,2}\big]~,
\> 
where $\kappa_{s,c}$ is the same prescription dependent constant we introduced above. 

These results can be extended to the generalised LL-model by including the higher-loop gauge theory  corrections to the interaction vertex and propagator. One finds, at tree-level in the two-dimensional theory, to ${\cal O}(g^2)$ that
 \<
 \left. F^{0,\Op{3}}_{s,c}(p_1,p_2,p_3)\right|_{{\cal O}(g^2)}&=& \frac{8(4 b_2 +5b_1) \kappa_{s,c}}{b_0 ~p_{12}^2p_{23}^2 p_{31}^2} \big[ p_2^2 p_{31}^2\chi_{1,2,3}
 \nn \\
 & &\kern-50pt +p_1^2 p_{23}^2\chi_{3,1,2}+p_3^2 p_{12}^2\chi_{2,3,1}\big]~.
 \> 
 The result at ${\cal O}(g^4)$ is similarly computable but somewhat more complicated
  \<
 \left. F^{0,\Op{3}}_{s,c}(p_1,p_2,p_3)\right|_{{\cal O}(g^4)}&=& \frac{2-\kappa_{s,c}}{2b_0^2}(5 b_1(5b_1+4b_2)-b_0(35 b_3+6 b_4+b_5))p_1p_2p_3(p_1+p_2+p_3)\nn\\
\kern-10pt +\frac{ 16 \kappa_{s,c}}{b_0^2 p_{12}^2p_{23}^2 p_{31}^2} 
 \Big[& &\kern-25pt(10 b^2_1+8 b_1 b_2-b_0(21 b_3+2 b_4+ b_5))p_1^7(p_2^3-p_2 p_3(p_2+p_3)+p_3^3)\nn\\
 -& &\kern-25pt(5 b_1^2+4 b_1 b_2+b_0(-63 b_3+2 b_4-5 b_5))p_1^6(p_2^4+p_3^4)\nn\\
 -& &\kern-25pt\half (85b_1^2+68 b_1 b_2+b_0(49 b_3-30 b_4+11 b_5))p_1^6(p_2^3 p_3+p_3^3 p_2)\nn\\
 +& &\kern-25pt(95 b_1^2+76 b_1 b_2+b_0(-77 b_3-26 b_4+b_5))p_1^6 p_2^2 p_3^2\nn\\
 +& &\kern-25pt4(5b_1^2 +4 b_1 b_2 -2 b_0(14b_3+ b_5))p_1^5p_2^5\nn\\
 -& &\kern-25pt\half(15 b_1^2+12 b_1 b_2+b_0(-119b_3+2b_4-9b_5))p_1^5(p_2^4 p_3+p_3^4 p_2)\nn\\
 +& &\kern-25pt\half(-25b_1^2-20 b_1 b_2+b_0(105 b_3+ 2 b_4+7 b_5))p_1^5(p_2^3 p_3^2+p_3^3 p_2^2)\nn\\
 +& &\kern-25pt(35b_1^2+28 b_1 b_2-b_0(161 b_3+ 2 b_4+11 b_5))p_1^4 p_2^4p_3^2\nn\\
 +& &\kern-25pt(-15b_1^2-12 b_1 b_2+b_0(49 b_3+ 2 b_4+3 b_5))p_1^4 p_2^3 p_3^3\nn\\
 + & &\kern-25pt\rm{cylic~ permutations~ of~ particle~ indices}\Big]~.
 \> 
 
 \paragraph{Further Quartic Operators}
 Just as for the quadratic operators we can consider additional operators by distributing derivatives across the fields. There is now an even greater number of possibilities however we will still consider just a single case namely:
 \<
 \Op{4}=|\varphi|^2 (\varphi^\ast\acute{\varphi}-\varphi\acute{\varphi^\ast})~.
 \>
 As can be seen immediately at tree-level the two-particle diagonal form factor simply acquires an additional factor of $i(p_1+p_2)$. This is in fact the case also at loop order as when inserted in 
 a chain of bubbles, just as in  \figref{fig:2b2loop} but with the operator replacing an interaction vertex, the loop-momenta from the vertex contribution always cancel and the additional momentum factor can be pulled out. Hence we have 
  \<
 \left. F^{\Op{4}}(p_1,p_2)\right|_{{\cal O}(g^0)}&=& \frac{4i(p_1+p_2)}{\left(1-\frac{ip_1 p_2}{p_1-p_2}\right)^2}~.
 \> 
\section{Spin-Chain Results}
\label{sec:spin-chain}
For comparison with the Landau-Lifshitz calculation let us review some of 
the known results for the ${\cal N}=4$ SYM spin-chain/AdS$_5 \times$ S$^5$ string world-sheet theory. As described in the introduction, at weak coupling we can expand the ${\cal N}=4$ SYM $\mathfrak{su}(2)$-sector dilatation generator in powers of $g=\tfrac{\sqrt{\lambda}}{4\pi}$ \eqref{eq:dilexp}.
The one-loop part is essentially the 
Heisenberg XXX-spin chain Hamiltonian.
The eigenstates of the Hamiltonian are characterized by 
their particle-, or magnon-, number and the energy of a state is given as a sum over single-magnon energies. In terms of the magnon momentum, $p$, this is given by $\varepsilon(p)=4 \sin^2 \tfrac{p}{2}$. 
Introducing the usual rapidity variable $u(p)=\tfrac{1}{2}\cot \tfrac{p}{2}$ we have
 \<
 \label{eq:tree_eng}
\varepsilon(p)=\frac{1}{\tfrac{1}{4}+u(p)^2}~.
\>
The spin-chain S-matrix which describes the phase change as two-magnons are exchanged and can be used to construct the multi-particle wavefunctions is given by
 \<
 S_{\rm XXX}(p_1, p_2)=\frac{u(p_2)-u(p_1)+i}{u(p_2)-u(p_1)-i}~.
 \>
 
 \paragraph{Low-energy limit}
 As described, the Landau-Lifshitz theory can be found by considering the low-energy limit of spin-chain which can be done either at the level of the action or at the level of computed quantities. We recall here the corresponding limit for  the S-matrix and form factors. This can then be repeated for the higher-loop results where the spin-chain Hamiltonian is significantly more complicated or even unknown. 
   
 To take the LL-limit we rescale the magnon energy, $\varepsilon\to \kappa^2 \varepsilon$, and consider the small  $\kappa$ limit.  For the rapidity variable we have that to leading order  
 \<
 u(p)\simeq \frac{1}{\kappa \sqrt{\varepsilon}}=\frac{1}{p}~
 \>
 where the momentum is given by $p= \kappa \sqrt{\varepsilon}$ 
 and hence the S-matrix is 
 \<
 S_{\rm LL}(p_1,p_2)=\frac{\frac{1}{p_2}-\frac{1}{p_1}+i}{\frac{1}{p_2}-\frac{1}{p_1}-i}~.
 \>
 This is the quantum S-matrix for the LL-model and written in this fashion there is no small parameter. This is the S-matrix which is perturbatively computed by resumming all loop orders in the LL model. As can be seen,  while the momenta as not taken to be small it does not reproduce the complete spin-chain S-matrix.  To extract just the tree-level result we reintroduce the small parameter by rescaling the momenta 
 $p_i\to \gamma p_i$ and then take the small $\gamma$ limit so that
 \<
  S_{\rm LL}(p_1,p_2)=1+\sum_{i=1}^{\infty} \gamma^{i+1}T^{(i)}(p_1,p_2)
  \>
  with 
 \<
 T^{(0)}_{\rm LL}( p_1,p_2)=\frac{2i p_1 p_2}{p_1-p_2}~,
 \> 
which is the same as the leading term in the tree-level T-matrix computed perturbatively \eqref{eq:T0gLL}.

\paragraph{The higher-loop spin-chain }
The extension to higher-loops in $g=\tfrac{\sqrt{\lambda}}{4\pi}$ can be described
in terms of the generalised $u(p)$ functions
\<
u(p)=\frac{1}{2}\cot \frac{p}{2}\sqrt{1+16 g^2 \sin^2 \frac{p}{2}}~.
\>
The all-order magnon energy is given by 
\<
2 g^2 \varepsilon(p)=\sqrt{1+16 g^2 \sin^2 \frac{p}{2}}-1
\>
and the S-matrix is 
\<
 S_{\rm {\cal N}=4}(p_1, p_2)=\frac{u(p_2)-u(p_1)+i}{u(p_2)-u(p_1)-i}\sigma(u_1,u_2)^2
 \>
 where $\sigma(u_1, u_2)$ gives the well-known dressing phase. As the dressing phase doesn't contribute until $\order(g^8)$ it can be ignored for our purposes. 
 
To study the low-energy limit to three-loops we again rescale $\varepsilon\to \kappa^2 \varepsilon$ but additionally 
we define $\tilde g=\kappa g$ which is essentially the effective coupling that appears in the BMN and other fast string expansions. We expand the magnon energy to ${\cal O}(\tilde g^4)$ so that
\<
\kappa^2 \varepsilon =4 \sin^2 \tfrac{p}{2}-16 \frac{\tilde g^2}{\kappa^2}\sin^4 \frac{p}{2}+-128 \frac{\tilde g^4}{\kappa^4}\sin^4 \frac{p}{2}~.
\>
In the limit of small $\kappa$ this implies
\<
\label{eq:mom_scale}
p=\kappa \sqrt{\varepsilon}\left(1+\tilde g^2 \frac{\varepsilon}{2}-\tilde g^4\frac{\varepsilon^2}{8} \right)
\>
or $\varepsilon=\tilde p^2-\tilde g^2 \tilde p^4+2 \tilde g^4 \tilde p^6$ where $\tilde p= \tfrac{p}{\kappa}$. Taking the same limit for
$u(p)$ we find 
\<
u(p)=\frac{1}{\kappa}\tilde u(\tilde p)=\frac{1}{\kappa}\left(\frac{1}{\tilde p}+2 \tilde g^2 \tilde p-2 \tilde g^4 \tilde p^3 \right)
\>
so that 
\<
 S_{\rm gen-LL}(p_1, p_2)&\equiv&
\frac{\tilde u(p_2)-\tilde u(p_1)+i}{\tilde u(p_2)-\tilde u(p_1)-i}
\nn\\
&=& \frac{1+\frac{i  p_1 p_2}{p_1-p_2}(1+2 \tilde g^2 p_1 p_2-2\tilde g^4 p_1 p_2(p_1^2-p_1 p_2 +p_2^2))}{1-
\frac{i  p_1 p_2}{p_1-p_2}(1+2 \tilde g^2 p_1 p_2-2\tilde g^4 p_1 p_2(p_1^2-p_1 p_2 +p_2^2))}~.
 \>
 This is the quantum S-matrix for the generalised LL-model. As in the LL-model, 
in order to define the perturbative two-dimensional expansion 
we again rescale the momenta $p_i\to \gamma p_i$ however
in order to keep the correct scaling result we write\footnote{This careless use of notation gives sensible results as we naturally think of both small parameters corresponding to the same large volume expansion, $\kappa \simeq \gamma \simeq \tfrac{1}{L}$.} $\tilde g= \tfrac{g}{\gamma}$ so 
that in the small $\gamma$ limit we have 
\<
 T^{(n)}_{\rm gen-LL}(p_1, p_2)=2\Big[\frac{ i p_1 p_2}{(p_1-p_2)}(1+2  g^2 p_1 p_2-2  g^4 p_1 p_2(p_1^2-p_1 p_2 +p_2^2))\Big]^{n+1}
  \>
  This result can be compared with the perturbative results above, \eqref{eq:T0gLL} and \eqref{eq:T1spin}, and it can be seen that they agree. 
  
 \subsection{Form Factors from XXX Spin-chain}
 \label{sec:XXXFF}
 Infinite volume diagonal spin-chain form factors, $f^{\cal{ O}}(p_1, \dots, p_n)$, have been calculated in \cite{Hollo:2015cda} by extracting them from 
 finite volume matrix elements. We will compare the low-energy limit of these results with those calculated directly from  the LL-model and then consider the generalisation to higher orders in $g$. 
 
\subsubsection{Length-one Operators}
 \label{sec:L1XXXFF}
The length-one operators, which correspond to the gauge theory operators ${\rm Tr}(Z\bar Z)$, 
${\rm Tr}(Z\bar X)$, ${\rm Tr}(X\bar Z)$ and ${\rm Tr}(X\bar X)$, are described by
the spin-chain operators acting on the $n$-th spin-chain site:
\<
E^{11}_n=\frac{1}{2}(\unit+\sigma^z_n)~, ~~~E^{12}_n=\sigma^+_n~, ~~~E^{21}_n=\sigma^-_n~, ~~~E^{22}_n=\frac{1}{2}(\unit-\sigma^z_n)~.
\>
For example, denoting $o_1(n)=E^{11}_n$, the vacuum, one-particle and two-particle diagonal form factors computed in \cite{Hollo:2015cda} were
\<
\label{eq:o1ffsc}
f^{o_1}(\emptyset)=1~,~~~f^{o_1}(p)=\epsilon(p)~,~~~f^{o_1}(p_1, p_2)=(\varepsilon(p_1) + \varepsilon(p_2))\phi_{12}
\>
where $\varepsilon(p)$ is the magnon energy as above and 
\<
\label{eq:phi_tree}
\phi_{12}=\frac{2}{1+(u(p_1)-u(p_2))^2}~.
\>
These can be compared to the previous perturbative results by using the map between the spin-chain and the LL-model via the coherent state representation, \eqref{eq:CSmap} and \eqref{eq:varphi}, whereby the spin-chain operator $o_1$ corresponds to the LL operator
\<
\label{eq:spintoLL}
o_1=\tfrac{1}{2}(1+\sigma_n^z) \leftrightarrow  1-|\varphi|^2~.
\>
We can see that compared to the LL operator considered in \secref{sec:length1op} there is an additional identity operator. This contribution gives rise to the non-trivial vacuum expectation value but can be ignored for higher-particle form factors.
To extract the prediction for the LL-model we must also perform the low-energy rescaling described above i.e. $\varepsilon\to \kappa^2 \varepsilon$ with $\kappa \to 0$. However in this limit all the multi-particle form factors \eqref{eq:o1ffsc} will vanish due to the normalisation of the one-particle states which results in the factors of the magnon energy. To get a well defined limit we rescale by a factor of $\sqrt{\varepsilon(p_i)}$ for each external leg. This results in the one-particle form factor being equal to $1$ which corresponds to the normalisation used in the perturbative calculation. For the two-particle case we find after this rescaling
\<
\label{eq:sc2parto1}
\frac{f^{o_1}(p_1, p_2)}{\varepsilon_1\varepsilon_2}\to \left(\frac{1}{\varepsilon(p_1)}+\frac{1}{\varepsilon(p_2)}\right)\frac{4}{1+\left(\frac{1}{p_1}-
\frac{1}{p_2}\right)^2}
\>
where on the r.h.s. we understand the dispersion relation to be that of the LL-model i.e. $\varepsilon(p)=p^2$. As a rule of thumb we see that the LL limit of infinite volume spin-chain quantities is taken by replacing 
$u(p)\to \tfrac{1}{p}$
while keeping constant terms that occur with differences of $u's$. For example in the quantities $\phi_{ij}$ we have
\<
\label{eq:phi_LL}
\phi_{ij}\to \phi^{\rm LL}_{ij}=\frac{2p_i^2 p_j^2}{(p_i-p_j)^2}\Big[1+\frac{p_i^2p_j^2}{(p_i-p_j)^2}\Big]^{-1}~,
\>
where the terms that arise in the small momentum expansion corresponds to world-sheet loop effects in the LL-model.
For the two-particle form factor it is apparent that this result \eqref{eq:sc2parto1} still does not match the LL result \eqref{eq:LL2parto1}. However this is again a consequence of the definition of the states used in defining the form factor and in fact 
\<
\frac{1}{\varepsilon_1\varepsilon_2} S_{\rm LL}(p_1,p_2) f^{-o_1}(p_1, p_2)\to F_s^{|\varphi|^2}(p_1,p_2)
\>
with the $S$-matrix factor due to differing ordering of momenta in the in- and out-states. 

A formula for multi-particle form factors was also proposed in \cite{Hollo:2015cda}.  For the length-one operator $o_1$
\<
\label{eq:o1ff}
f^{o_1}(p_1,\dots, p_n)=\sum_{\sigma\in S_n}\varepsilon_{\sigma(1)} \phi_{\sigma(1)\sigma(2)}\phi_{\sigma(2)\sigma(3)}\dots\phi_{\sigma(n-1),\sigma(n)}~,
\>
where the sum is over the set of all permutations of the $n$-indices, $S_n$. 
As in the two-particle case in order to have a non-vanishing answer in the LL-limit we must rescale 
by a factor of $(\varepsilon_1 \dots \varepsilon_n)^{-1}$ and thus taking the limit we find
\<
\frac{f^{o_1}(p_1,\dots, p_n)}{\varepsilon_1\dots \varepsilon} \to & &\\
& &\kern-100pt 
2^{n-1}\sum_{\sigma\in S_n}
\frac{ p^2_{\sigma(1)} \dots p^2_{\sigma(n-1)} }{ (p_{\sigma(1)}-p_{\sigma(2)})^2\dots (p_{\sigma(n-1)}-p_{\sigma(n)})^2 }
\frac{1}{
	1+\left(\tfrac{p_{\sigma(1)} p_{\sigma(2)} }{ p_{\sigma(1)}-p_{\sigma(2)}}  \right)^2} \dots 
\frac{1}{
	1+\left(\tfrac{p_{\sigma(n-1)} p_{\sigma(n)} }{ p_{\sigma(n-1)}-p_{\sigma(n)} } \right)^2}~.\nn
\>
To compare with the tree-level Landau-Lifshitz result for three particles computed in \secref{sec:length1op} we expand in powers of the momenta and take the leading results. Up to an overall sign agreement is found. At tree-level the S-matrix is simply $1$, however we would expect to see factors of the S-matrix by keeping higher orders in the momenta corresponding to loop effects in the perturbative calculation.

 \paragraph{Higher-loop Form Factors}
As seen in \secref{sec:LLff}, one can straightforwardly calculate higher-loop form factors in the generalised Landau-Lifshitz model.
An $\order(g^2)$ prediction for length-one form factors was given in \cite{Jiang:2015bvm} where they were related to the computation of certain structure constants. 
The explicit perturbative computation involves several contributions: corrections to the states due to the two-loop gauge theory corrections to the dilatation generator and
modifications due to operator insertions capturing the effects of one-loop gauge theory Feynman diagrams \cite{Okuyama:2004bd, Alday:2005nd, Petrovskii:2016sfs}.
As a result, in the form factor picture the operator itself must be viewed as acquiring $\order(g^2)$ corrections\footnote{We thank Y. Jiang for emphasising this point to us.}
\<
o_1(g)=o_1+g^2 o_1'~.
\>
Somewhat remarkably, the ``sum over products" form of the tree-level result \eqref{eq:o1ff} remains, with the corrections coming in the individual  
components. Specifically
\<
f^{o_1(g)}(u_1, \dots, u_n)=\sigma_1 \varphi_{12}\varphi_{23}\dots \varphi_{n-1,n}+{\rm permutations}
\>
where
\<
\sigma_i=\frac{1}{u_i^2+\tfrac{1}{4}}+\frac{8g^2 u_i^2}{(u_i^2 +\tfrac{1}{4})^3}
 \>
 and 
 \<
 \varphi_{ij}=\frac{2}{(u_i-u_j)^2+1}+\frac{4 g^2 (u_i^2-u_j^2)}{(u_i^2+\tfrac{1}{4})(u_j^2+\tfrac{1}{4})((u_i-u_j)^2+1)}.
\>

For the one-particle form factor, $f^{o_1(g)}(u(p_1))$, at the leading order, ${\cal O}(g^0)$, we rescaled by the energy of the external particle to find agreement with the perturbative LL calculation. As the $\sigma_i$'s do not correspond to the $g$-corrected expression for the particle energy which is instead given by 
 \<
 \epsilon(u)=\frac{1}{u^2+\tfrac{1}{4}}+g^2\frac{12 u^2-1}{4(u^2+\tfrac{1}{4})^3}~, 
\>
we must add a correction to the operator. Computing the small momentum limit we have
\<
\frac{1}{\epsilon}f^{o_1(g)}(u)\to 1+5g^2 p^2
\>
hence by considering the generalised LL operator $\Phi_1(g)=\varphi^\ast \varphi+5 g^2\acute \varphi \acute{\varphi}^\ast$ we have that
\<
\frac{1}{\epsilon}f^{o_1(g)}(u)\to F^{\Phi_1(g)}(p)~.
\>
For the two-particle form factor, again dividing by factors of the particle energy and expanding in powers of the momenta, we have
\<
\frac{1}{\varepsilon_1 \varepsilon_2}f^{o_1(g)}(p_1,p_2) \to \frac{2(p_1^2+p_2)^2}{(p_1-p_2)^2}\left(1+g^2 (3p_1^2+4 p_1 p_2+3 p_2^2)\right)~.
\>
This can be seen to not agree with \eqref{eq:ff2c} and also does not reproduce $F^{\Phi_1(g)}(p_1,p_2)$ when the coefficients $b_1$ and $b_2$ appearing in the generalised LL action are set to their string/BDS value. One can include further corrections to the operator, which as long at they are at least quartic in the fields won't change the one-particle form factor result,
\<
\Phi_1(g)=\varphi^\ast \varphi+5 g^2\acute \varphi \acute{\varphi}^\ast+ \alpha_1 g^2 (\varphi^\ast{}^2 \acute{\varphi}^2+\varphi^2\acute{\varphi}^\ast{}^2)+
\alpha_2 g^2 \varphi \varphi^\ast \acute{\varphi} \acute{\varphi}^\ast,
\>
however there do not appear to be values of $\alpha_1$ and $\alpha_2$ that correctly reproduce the limit of the two-particle form factor of $o_1(g)$ and it seems that a more general deformation or extra contribution appears to be required. \footnote{In \cite{Petrovskii:2016sfs} an operator correction reproducing the effect of the insertions to the heavy operator for the one-particle form factor constructed, responsible for the ``$\delta_H$" correction, was given. It corresponds to $\Phi'_1=\varphi^\ast \varphi+2 g^2\acute \varphi \acute{\varphi}^\ast+  g^2 (\varphi^\ast{}^2 \acute{\varphi}^2+\varphi^2\acute{\varphi}^\ast{}^2)$. However this doesn't reproduce the full two-particle form factor.}
\subsubsection{Length-two Operators}
\label{sec:l2op}
We can additionally consider the length-two spin-chain operators
\< 
o^1_{2}=E_n^{11}E_{n+1}^{11}~,~~~o_2^2=E_n^{12}E_{n+1}^{21}~,~~~o_2^3=E_n^{21}E_{n+1}^{12}
\>
where the operators now sit on two spin-chain lattice sites and  
the infinite volume form factors were again extracted from spin-chain matrix elements in 
\cite{Hollo:2015cda}. For each operator, ${\cal O}\in\{o^1_{2},o^2_{2},o^3_{2}\}$, they can be written as a combination of two terms 
\<
\label{eq:scff}
f^{\cal O}(u_1, \dots, u_n)&=&f_E^{\cal O}(u_1, \dots, u_n)+f_S^{\cal O}(u_1, \dots, u_n)
\>
with each term given as a sum over permutations
\<
f_E^{\cal O}(u_1, \dots, u_n)=\sum_{\sigma\in S_n} \Big[\varepsilon_{\sigma(1)} \phi_{\sigma(1)\sigma(2)}\dots
\phi_{\sigma(n-1)\sigma(n)}\mathfrak{f}^{\cal O}_n\Big]
\>
and
\<
f_S^{\cal O}(u_1, \dots, u_n)=\sum_{\sigma\in S_n}\Big[ \sum_{i=1}^{n-1} \varepsilon_{\sigma(1)} \phi_{\sigma(1)\sigma(2)}\dots\psi^{\cal O}_{\sigma(i-1)\sigma(i)}\dots 
\phi_{\sigma(n-1)\sigma(n)}\varepsilon'_{\sigma(n)}\Big]
\>
where the energy, $\varepsilon$ is as in \eqref{eq:tree_eng}, $\varepsilon'$ is the derivative of 
the energy with respect to the rapidity variable $u$,  $\phi_{ij}$ is as in \eqref{eq:phi_tree}
and 
\<
	\renewcommand*{\arraystretch}{1.5}
\begin{array}{ll}
\mathfrak{f}^{o_2^1}_i=2~, &~~~\psi^{o^1_2}_{ij}=-(u_i-u_j)(u_iu_j-\frac{1}{4})\phi_{ij}~,\\
\mathfrak{f}^{o_2^2}_i=\frac{u_i-i/2}{u_i+i/2}~, & ~~~\psi^{o^2_2}_{ij}=(u_i-u_j)(u_i-i/2)(u_j-i/2)\phi_{ij}~,\\
 \mathfrak{f}^{o^3_2}_{i}=\frac{u_i+i/2}{u_i-i/2}~, & ~~~\psi^{o^3_2}_{ij}=(u_i-u_j)(u_i+i/2)(u_j+i/2)\phi_{ij}~.\\
\end{array}
\>
Of course one can consider linear combinations of these operators and one such combination 
in which we will be interested is
\<
o_2^4=E_n^{22}E_{n+1}^{22}=\unit-o_1(n)-o_1(n+1)+o_2^1(n)
\>
and for which we have
\<
\label{eq:o24ff}
\mathfrak{f}^{o_2^4}_i=0~, & & ~~~\psi^{o^4_2}_{ij}=-(u_i-u_j)(u_i u_j-\frac{1}{4})\phi_{ij}~.
\>
Taking the continuum limit by using the replacement rule \eqref{eq:spintoLL} it is easy to see that
the operator $o_2^4$ corresponds to the length-two LL operator $|\varphi|^4$ up to 
derivative terms which we neglect. 
Thus we can compare the form factors for this operator with those previously calculated perturbatively. The tree-level results can be found by simply making a small momentum expansion. Explicitly, this gives 
\<
\varepsilon'\to -2p^3~,~~~{\rm and}~~~\psi^{o^1_4}_{ij}\to \frac{2}{(p_i-p_j)}~,
\>
in addition to $\varepsilon\to p^2$ and $\phi_{ij}\to\phi^{\rm LL}_{ij}$.
It is easy to see that the two-particle form factor, once rescaled, has in the small momemtum limit the trivial result
\<
\frac{1}{\varepsilon_1 \varepsilon_2}f^{o_2^4}(p_1,p_2)\to \left. F_c^{(0)|\varphi|^4}(p_1,p_2)\right|_{{\cal O}(g^0)}=4
\>
while the three-particle case gives
\<
\frac{1}{\varepsilon_1 \varepsilon_2 \varepsilon_3}f^{o_2^4}(p_1,p_2,p_3)\to \left. -F_c^{(0)|\varphi|^4}(p_1,p_2,p_3)\right|_{{\cal O}(g^0)}
\>
which means it agrees with the tree-level LL result up to a sign.

Furthermore the loop effects are reproduced by using our rule of thumb of
replacing $u\to \tfrac{1}{p}$ and keeping those constants that are added to differences of $u$'s. In particular this does not retain the factor of $\tfrac{1}{4}$ in the definition of $\psi_{ij}^{o^4_2}$ . These factors can be reproduced in the LL model by adding derivative terms to the operator. 

\section{Form Factor Perturbation Theory}
One interesting application of diagonal form factors is to the study of perturbations of integrable models. Such an approach, form factor perturbation theory (FFPT), to studying non-integrable massive theories was introduced in \cite{Delfino:1996xp} with a particular focus on deformations of relativistic integrable models which themselves can be viewed as deformations of conformal field theories. However, as the authors of \cite{Delfino:1996xp} make clear, their approach is quite general. Given an integrable model with action ${\cal A}_0^{\rm int}$ they study a theory with an action
\<
{\cal A}={\cal A}_0^{\rm int}-\sum_j g_j\int d^2 x~ \Phi_j(x)
\>
where $\Phi_j(x)$ are the deforming operators. An assumption behind this approach is that, at least for small values of $g_j$, asymptotic particle states are a good basis for studying the deformed theory and that while the integrable model has a different spectrum it acts as a useful starting point. We will be interested in calculating the S-matrix of the deformed theory
\<
\label{eq:Smatdef}
S(p_1, \dots, p_n;k_1, \dots, k_m)={}^{\rm out}\langle k_1, \dots, k_m|p_1,\dots,p_n\rangle^{\rm in}~.
\>
In order to preserve the normalisation of the vacuum, in \cite{Delfino:1996xp}, the authors introduced a counter-term corresponding to the vacuum energy
so that
\<
\label{eq:vac}
 {}^{\rm out}\langle 0|0\rangle^{\rm in}=\leftidx{_{ \hphantom{00}0}^{\rm out}}{}{}\langle 0|0\rangle^{\rm in}_0=1
\>
where, for example,  $|0\rangle^{\rm in}_0$ is the ``in"-vacuum state in the undeformed theory.  They further introduced counter-terms to preserve the one-particle normalisation. Here we define the operators ${\cal O}^{(i)}(0,0)$, $i=1,2$,  in terms of their form factors in the unperturbed, integrable theory,
 \<
F^{{\cal O}^{(1)}}(p_1,\dots,p_n)=\delta_{n,1}~, ~~~{\rm and}~~~
	F^{{\cal O}^{(2)}}(p_1,\dots,p_n)=i p_1 \delta_{n,1}
\>
such that\footnote{As we will be considering the Landau-Lifshitz model which is not Lorentz invariant we
modify several of the definitions of \cite{Delfino:1996xp}; for example we don't use the usual Lorentz invariant one-particle normalisations.}
\<
\label{eq:one_part}
{}^{\rm out}\langle k|p \rangle^{\rm in}= \leftidx{_{ \hphantom{00}0}^{\rm out}}{}{}\langle k|p \rangle^{\rm in}_0=2\pi \delta(p-k)~.
\>
As described in \cite{Delfino:1996xp}, a perturbative expansion for the S-matrix can be found by expanding \eqref{eq:Smatdef} in terms of matrix elements of time-ordered products of the deformations and inserting sums over asymptotic states of the undeformed theory. In principle this gives an expansion to all orders in the couplings $g_j$, with higher orders involving progressively more sums over intermediate states much as in covariant perturbation theory. Here we will only consider the leading-order terms
\<
\label{eq:LinDef}
{}^{\rm out}\langle k_1, \dots, k_m|p_1,\dots,p_n\rangle^{\rm in}&\simeq&
\leftidx{_{ \hphantom{00}0}^{\rm out}}{}{}\langle k_1, \dots, k_m|p_1,\dots,p_n\rangle^{\rm in}_0\\
& &\kern-150pt - i (2\pi)^2 \delta^{(2)}(\sum k_i-\sum p_j) 
\leftidx{_{ \hphantom{00}0}^{\rm out}}{}{}\langle k_1, \dots, k_m|\Big[\sum_j g_j \Phi_j-\sum_i\delta{\cal E}^{(i)}{\cal O}^{(i)}-\delta {\cal E}_{\rm vac}\Big]|p_1,\dots,p_n\rangle^{\rm in}_0\nn
\>
where the coefficients $\delta{\cal E}^{(i)}$ and $\delta {\cal E}_{\rm vac}$ are determined by demanding that 
\eqref{eq:vac} and \eqref{eq:one_part} are satisfied.

\subsection{Marginal Deformations}
The use of integrable models in the study of the AdS/CFT correspondence has been very fruitful but the vast majority of interesting theories are almost certainly non-integrable. The corresponding world-sheet theories will likely involve multi-particle production with a corresponding increase in analytical complexity of the world-sheet S-matrix. Leigh-Strassler marginal deformations are one particularly simple class of deformations of ${\cal N}=4$ SYM that preserve ${\cal N}=1$ superconformal symmetry \cite{Leigh:1995ep} and which are parameterised by two complex parameters $h$ and $q={\rm exp}(2\pi i \beta)$; thus, including the gauge coupling, there is a three-dimensional space of finite theories. The case with $h=0$ is often called the $\beta$-deformed theory and, particularly for real $\beta$, it has received a very significant amount of attention as the gravitational dual is known \cite{Lunin:2005jy} and the model is believed to be integrable -- the string Lax pair was constructed in \cite{Frolov:2005dj}, the all-loop asymptotic Bethe ansatz was proposed in \cite{Beisert:2005if} and Y-system in \cite{Gromov:2010dy}. The integrability of the $\beta$-deformed theory can be understood as arising from a Drinfeld-Reshetikhin twist of the undeformed theory combined with twisted boundary conditions \cite{Ahn:2010ws, Ahn:2011xq}. 

For complex $\beta$, the one-loop dilatation operator restricted to two holomorphic scalar fields corresponds to the $\mathfrak{su}(2)_q$ XXZ deformed spin chain and so is integrable \cite{Roiban:2003dw}, however this does not extend beyond this subsector of fields \cite{Berenstein:2004ys}. For special values of $h \neq 0$ and $q \in \mathbbm{C}$ the one-loop Hamiltonian is integrable \cite{Bundzik:2005zg} which can be understood in terms of Hopf twists of the real-$\beta$ case \cite{Mansson:2008xv}. More generally for generic values of $q$ and $h$ the theory is not believed to be integrable. For general $q$ and $h$ the R-matrix constructed by applying the Hopf algebraic transformation will not satisfy the Yang-Baxter equations and so the usual methods of integrable spin-chains will not be applicable, such deformations may however be studied by use of FFPT. In some sense the one-loop marginal deformations in the $\mathfrak{su}(2)$ sector which we study below are too simple to be of much interest, however they will allow us the check the general formula against known results and so demonstrate its reliability to this order. 

\subsection{Deformed Landau-Lifshitz}
Here will consider the Landau-Lifshitz model following from the low-energy limit of the general Leigh-Strassler deformed one-loop spin-chain given in \cite{Bundzik:2005zg}. The LL model for complex-$\beta$ but $h=0$ was considered in \cite{Frolov:2005ty}\footnote{In \cite{Frolov:2005ty} the authors use the complex parameter $\beta_\mathbbm{C}=\beta_d+i \kappa_d$ where $\beta_d=\frac{\beta_{R}}{2\pi}$ and $\kappa_d=\frac{\beta_{I}}{2\pi}$.} where checks of the match between the spin-chain and string descriptions were carried out. In keeping with our previous considerations we will truncate to the case of two holomorphic scalars such that the spin-chain Hamiltonian is given by 
\<
\label{eq:def_ham}
H^D&=&\frac{\lambda}{16\pi|q|}\sum_{\ell=1}^L\Big[(\frac{1+q q^\ast}{2}+h h^\ast)\unit\otimes \unit-(\frac{1+q q^\ast}{2}-h h^\ast)\sigma^z_\ell\otimes \sigma^z_{\ell+1}\nn\\
& &\kern+50pt -2q~ \sigma^-_\ell\otimes \sigma^+_{\ell+1}-2q^\ast \sigma_\ell^+\otimes \sigma_{\ell+1}^-\Big]~.
\>
Using the parameterisation $q={\rm exp}(\beta_I  +i\beta_R)$, $2 h h^\ast e^{-\beta_I}=\Delta^2$ and taking the Landau-Lifshitz limit we find that in order to have a sensible behaviour the deformation parameters must be taken to be small with $\tilde{\beta}_{R}=\tfrac{\beta_R L}{2\pi}$, $\tilde{\beta}_{I}=\tfrac{\beta_I L}{2\pi}$ and $\tilde \Delta=\tfrac{L\Delta}{2\pi}$ fixed.  With this scaling the resulting deformed Landau-Lifshitz action is
\<
{\cal A}&=&{\cal A}^{\rm LL}-\frac{\lambda}{16\pi L}\int d\tau d\sigma \Big[ \tilde{\beta}_{R}^2 \left((n^1)^2+(n^2)^2\right)+2 \tilde{\beta}_{R}\left(n^1 \acute{n}^2-n^2\acute{n}^1\right)
\\
& &\kern+120pt+ \tilde{\beta_I}^2 \left(1-(n^3)^2\right)+\tilde{\Delta}^2 \left(1+(n^3)^2\right)\Big]\nn
\>
where ${\cal A}^{\rm LL}$ is the Landau-Lifshitz action \eqref{eq:LLaction}. Setting $\tilde \Delta=0$ and 
using
\<
(n^1,n^2,n^3)=(\sin 2\theta \cos 2\eta, \sin 2\theta \sin 2 \eta,\cos 2\theta)
\>
one reproduces the result from \cite{Frolov:2005ty}. Instead we rescale the coordinates as in \eqref{eq:xtcoord} so that the spatial coordinate has period $L$, use the complex field $\varphi$ defined in \eqref{eq:varphi} and expand the action to quartic powers in the field 
\<
{\cal A}&=&{\cal A}^{\rm LL}-
\int dxdt~ \Big[\frac{\Delta^2}{2}  +(\beta_I^2-\Delta^2)\Phi_1-i\beta_R \Phi_2\nn\\
& &\kern+120pt+(\Delta^2-\beta_I^2)\Phi_3+ i\beta_R \Phi_4
\Big]
\>
where the deformations are given by the operators considered previously
\<
\Phi_1=|\varphi|^2~,~~~\Phi_2=(\varphi^\ast \acute{\varphi}-\varphi \acute{\varphi}^\ast)~,~~~
\Phi_3=|\varphi|^4~,~~~\Phi_4=|\varphi|^2(\varphi^\ast\acute{\varphi}-\varphi\acute{\varphi}^\ast)~.
\>
We will use the form factor perturbation procedure to describe the corrections to the S-matrix due to these deformations.

\paragraph{Integrable deformations} As the Hamiltonian \eqref{eq:def_ham} with $\Delta=0$ is in fact integrable the full Bethe equations are known and we will be able to compare our results with those previously calculated \cite{Frolov:2005ty},
\<
\label{eq:Def_BA}
e^{- i \beta_R L} \Big[\frac{ \tilde u_k+i/2}{ \tilde u_k -i/2}\Big]^L=\prod_{\stackrel{j=1}{j\neq k}}^M
\frac{ \tilde u_k-\tilde u_j+{i} \frac{\tanh\beta_I}{2\tanh\tfrac{\beta_I}{2}}(1+4 \tanh^2 \tfrac{\beta_I}{2} \tilde u_k \tilde u_j)}{ \tilde u_k-\tilde u_j-{i} \frac{\tanh\beta_I}{2\tanh\tfrac{\beta_I}{2}}(1+4 \tanh^2 \tfrac{\beta_I}{2} \tilde u_k \tilde u_j)}
\>
where
\<
e^{- i \beta_R M} \prod_{{k=1}}^M \frac{ \tilde u_k+i/2}{ \tilde u_k -i/2}=1
\>
and 
\<
E=\frac{\lambda}{8\pi^2}\sum_{j=1}^M \tilde \varepsilon_j~~~{\rm with}~~~\tilde \varepsilon_j=\frac{1}{\tilde u_k^2+1/4}+2(\cosh \beta_I-1)~.
\>
These equations give the corrections to the one-particle states, the two-particle S-matrix and the general $n$-particle S-matrix which can be found as a  product of two-particle S-matrices. The results calculated using form factor perturbation theory will be expressed in terms of rapidities and momenta of the undeformed theory. These can be related to the deformed rapidities using the relation
\<
\frac{ \tilde u+i/2}{ \tilde u-i/2}= e^{i \beta_R} \frac{ u+i/2}{  u-i/2}
\>
or to leading order in $\beta_R$, $\tilde u= u-( u^2 +1/4) \beta_R$. Hence we find the correction to the energy $\tilde \varepsilon=\varepsilon+\delta\varepsilon$, with 
\<
\label{eq:def_eng}
\delta \varepsilon(u)=\beta_R \frac{2 u}{ u^2+1/4}+\beta_I^2 
\>
and to the S-matrix
\<
\label{eq:def_Smat}
\delta S(u_1, u_2)=2i\beta_R \frac{  u_2^2 -u_1^2}{( u_1- u_2+i)^2}+i \beta_I^2 \frac{(u_1-u_2)(1-4  u_1 u_2)}{2( u_1- u_2+i)^2}~.
\>
We can take the low-energy limit of these results to compare with those calculated in the LL-model. The modification of the periodicity condition can be accounted for by shifting the relation between the rapidity and particle momentum
\<
\tilde u(p)=\frac{1}{2}\cot \frac{p+\beta_R}{2}~.
\>
To take the low-energy LL limit we take the momentum and $\beta_R$ to scale as $\kappa$ as $\kappa \to 0$,
so that we have 
\<
\tilde u(p)=\frac{1}{p+\beta_R}\simeq \frac{1}{p}-\frac{\beta_R}{p^2}+{\cal O}(\beta_R^2)~.
\>
The corresponding equation for the change in the energy is
\<
\label{eq:def_engy_int}
\delta \varepsilon=2\beta_R~ p+\beta_I^2
\>
and for the change in the S-matrix
\<
\label{eq:def_Smat_int}
\delta S(p_1,p_2)=\frac{  2i \beta_R (p_1+p_2)}{(p_1-p_2)\left(1-\frac{ip_1p_2}{p_1-p_2}\right)^2}+ \frac{ 2 i \beta_I^2(1+\gamma~ p_1 p_2 ) }{(p_1-p_2)\left(1-\frac{ip_1p_2}{p_1-p_2}\right)^2}~.
\>
Here we have introduced a parameter $\gamma$ in the $\beta_I^2$ deformations; in taking the LL-limit previously, \secref{sec:l2op},  we have kept sub-leading terms of the form $1/(u_1-u_2)$ but dropped those of the form $1/(u_1 u_2)$ which corresponds to setting $\gamma=0$.

\subsubsection{Deformed Landau-Lifshitz from Form Factor Perturbations}
We can use our previous perturbative calculations of the LL form factors in \secref{sec:LLff} and the general expression \eqref{eq:LinDef} to calculate the corrections to the S-matrix elements to linear order in the deformations. 

\paragraph{Vacuum Energy} As none of the operators $\Op{i}$ have non-vanishing zero-particle form factors the only correction to the vacuum energy comes from the coefficient of the identity operator, namely $\Delta^2$. Using the condition that the vacuum normalisation remains unchanged fixes the counterterm coefficient 
\<
\delta{\cal E }_{\rm vac}=\frac{\Delta^2}{2}~.
\>

\paragraph{One-particle states}
More interestingly, the quadratic operators $\Op{1}$ and $\Op{2}$ give rise to corrections to the one-particle state normalisations
\<
\delta{\cal E}^{(1)}=(\beta_I^2-\Delta^2)~,~~~\delta{\cal E}^{(2)}=-2i\beta_R~.
\>
These deformations correspond to corrections to the dispersion relation
\<
\omega(p)=p^2+2\beta_R~ p+(\beta_I^2-\Delta^2)
\>
 and to calculate the one-particle energies one should multiply by the factor of $\tfrac{\lambda}{8\pi^2}$ that arises from the rescaling of the time coordinate. If we consider the case $\Delta=0$ we have
\<
\varepsilon(p)=|p+\beta_{\mathbbm{C}}|^2
\>
which strictly speaking should only be trusted to $\order(\beta_R)$ in our calculations, where $\beta_{\mathbbm{C}}=\beta_R+i \beta_I$ and this result can be seen to agree with that previously calculated in the deformed Landau-Lifshitz model \cite{Frolov:2005ty}. 

\paragraph{Two-particle states}
For the two-particle form factors for the operators $\Phi_3$ and $\Phi_4$ we have the result
\<
\label{eq:S_var_1}
\delta S(p_1,p_2)&=&\frac{-i}{2(p_1-p_2)}\big[(\Delta^2-\beta_I^2)F^{\Phi_3}+i\beta_R F^{\Phi_4}\big]\nn\\
&= &  \frac{2i (p_1+p_2)\beta_R}{(p_1-p_2)\left(1-\frac{ip_1p_2}{p_1-p_2}\right)^2}
+ \frac{2i(\beta_I^2-\Delta^2 )}{(p_1-p_2)\left(1-\frac{ip_1p_2}{p_1-p_2}\right)^2}~. 
\>
were we have taken into account the Jacobian, $\tfrac{-i}{2(p_1-p_2)}$,  relating the usual energy-momentum $\delta$-function and the momentum $\delta$-functions in front of the S-matrix (in addition to the factor of $-i$ from \eqref{eq:LinDef}).
To compare with the Bethe Ansatz \eqref{eq:def_Smat_int} results we simply set $\Delta=0$ and the results can be seen to match. For the $\beta_R$ term this deformation essentially follows from the shift of the rapidities. 
The same deformed S-matrix could be found by taking the fast-string limit of the string world-sheet theory in the $\beta_R$-deformed geometry. A perturbative calculation \cite{Ahn:2012hs} of the world-sheet S-matrix in the near-BMN limit of the deformed theory  \cite{McLoughlin:2006cg} has been previously carried out and is consistent with the above result. For the $\beta_I^2$ term  we see that we find the $\gamma=0$ result. In order to reproduce the $\gamma=1$ result we would to have to add appropriate derivative corrections to the deformation operator.
 
In principle there should be additional corrections to the S-matrix from $\Phi_1$ and $\Phi_2$ which have non-vanishing two-particle diagonal form factors:
\<
\label{eq:S_var_2}
\delta S(p_1,p_2)&=&\frac{-i}{2(p_1-p_2)}\big[(\beta_I^2-\Delta^2)F^{\Phi_1}-i\beta_R F^{\Phi_2}\big]\nn\\
&=&\frac{2 i}{(p_1-p_2)^3\left(1-\frac{i p_1p_2}{p_1-p_2}\right)}
\big[(\beta_I^2-\Delta^2) (p_1^2+p_2^2)-2\beta_R ~p_1p_2(p_1+p_2)\big]
\>
where it is important to note that we use the symmetric prescription to evaluate the diagonal form factors. These corrections correspond to the changes in the S-matrix as a result of changes in the dispersion relation.  The relativistic analogue of this was discussed in \cite{Delfino:1996xp}, where as the invariant 
\<
s=2m^2(1+\cosh\theta)
\>
is held constant under the deformation of a parameter, which we call $\delta g$,  the resulting change in the particle mass, $\delta m$, necessarily causes a shift of the rapidity $\delta \theta =-2\tfrac{\delta m}{m} \coth \theta/2$ and so the change in the S-matrix has two components
 \<
 \delta S(\theta)=\frac{\partial S(\theta)}{\partial \theta}\delta \theta+\left. \frac{\partial S(\theta, g)}{\partial g}\right|_{g=0}\delta g~.
\>
The generalised LL-model, being non-relativistic, doesn't satisfy the same relation but we can define an analogous variation due to changes in the particle momenta
\<
\label{eq:varS}
\delta S(p_1,p_2)=\frac{\partial S(p_1,p_2)}{\partial p_1}\delta p_1+\frac{\partial S(p_1,p_2)}{\partial p_1}\delta p_1
+\sum_i \left. \frac{\partial S(p_1,p_2, g_i)}{\partial g_i}\right|_{g_i=0}\delta g_i
\>
where in this case we are considering $g_i\in \{\beta_R, \beta_I^2, \Delta^2\}$. The variations w.r.t. the couplings give the terms calculated previously  \eqref{eq:S_var_1} while the first two terms should correspond to \eqref{eq:S_var_2}. This is clearest for the $\beta_I^2, \Delta^2$ deformations where if we demand that total incoming momentum and energy are unchanged by the deformation i.e.  $\delta \varepsilon_1+\delta \varepsilon_2=0$
we have that 
\<
2 \delta p_1 p_1+2 \delta p_2 p_2=-2(\beta_I^2-\Delta^2)~, ~~~\delta p_1+\delta p_2=0~.
\>
and solving for $\delta p_1$ and $\delta p_2$ and substituting into the first two terms of \eqref{eq:varS} we find the corresponding terms in \eqref{eq:S_var_2}. To reproduce the $\beta_R$ terms we must modify the variation  conditions such that 
\<
2 \delta p_1 p_1+2 \delta p_2 p_2=-4(p_1+p_2)\beta_R-2(\beta_I^2-\Delta^2)~, ~~~\delta p_1+\delta p_2=-2\beta_R~.
\>

\paragraph{Integrable form factors}
As the spin-chain form factors have been computed via the algebraic Bethe ansatz, our results are in fact generalisable to that theory without the need to take the LL low-energy limit. While the perturbative approach we have taken can be used to find the deformed S-matrix for low numbers of external particles, such integrable methods potentially give a method to  completely determine the $n$-particle S-matrix. To leading order in the deformations, we can write the Hamiltonian as 
\<
H^D=H^{\rm XXX}+\frac{\lambda}{8\pi }{\cal O}^D
\>
where 
\<
{\cal O}^D=\sum_{\ell=1}^L\Big[i \beta_R \left(o^2_2(\ell)-o^3_2(\ell)\right) +(\Delta^2-\beta_I^2)\left(o^1_2(\ell)-o^1(\ell)\right)+\tfrac{1}{2}\Delta^2\Big]~.
\>
A proposal for the $n$-particle diagonal form factors (including $n>2$) of this deformation can be given by simply taking linear combinations of the results found in \cite{Hollo:2015cda}, multiplying by the appropriate factor of the \textit{undeformed} S-matrix, and including factors of particle energies to correct the state normalisations:
\<
F^{{\cal O}_D}(\emptyset)=\tfrac{1}{2}\Delta^2~,~~~ ~~~F^{{\cal O}_D}(u_1)=\frac{1}{\epsilon_1}f^{{\cal O}_D}(u_1)
\>
and for $n\geq 2$
\<
F^{{\cal O}_D}(u_1, \dots,  u_n)=\frac{\prod_{i\neq j}^n S( u_i,  u_j)}{\epsilon_1\dots \epsilon_n} f^{{\cal O}^D}( u_1, \dots, u_n)
\>
where the spin-chain form factors $f^{{\cal O}^D}$ are given by \eqref{eq:scff} with
\<
\mathfrak{f}^{{\cal O}^D}&=&i \beta_R(\mathfrak{f}^{o^2_2}-\mathfrak{f}^{o^3_2})+(\Delta^2-\beta_I^2)(\mathfrak{f}^{o_2^1}-1)\nn\\
&=&-2 \beta_R \frac{u}{u^2+\tfrac{1}{4}}+(\Delta^2-\beta_I^2)
\>
and 
\<
\psi^{{\cal O}^D}_{ij}=( u_i- u_j)\Big[\beta_R ( u_i+ u_j)-(\Delta^2-\beta_I^2)( u_i  u_j-\tfrac{1}{4})\Big]\phi_{ij}~.
\>
These can now be used to compute the corrections to the spin-chain S-matrix. As can be seen by comparison with \eqref{eq:def_eng} the factor $\mathfrak{f}^{{\cal O}^D}$, which only contributes to the $f_E$ term in \eqref{eq:scff}, gives the (negative of) the corrections to the magnon energies which is consistent as it is the sole contribution to the $n=1$ form factor
in the absence of the counterterms. Similarly by comparison with \eqref{eq:def_Smat} we can see that the deformation of the two-particle S-matrix is reproduced entirely by the $f_S$ part of the two-particle form factor from \eqref{eq:scff} which gives
\<
\delta S=-i {S(u_1,u_2)\psi_{12}}~.
\>
with the additional terms appearing in $F^{{\cal O}_D}(u_1,u_2)$ being cancelled by the Jacobian from the energy-momentum $\delta$-functions. Importantly here we are not taking the low-energy LL-limit and the results are valid for arbitrary momenta in the infinite volume limit and in particular we capture the factor of $\tfrac{1}{4}$ in $\psi^{\order^D}_{12}$ that is missed in the LL limit. There is additionally a contribution from the $f_E$ part of the two-particle form factor; as in the LL theory these should be related to the change in the S-matrix due to the change in the definition of the rapidity.

\section{Outlook}

While there are a number of different directions to pursue - for example other form factors at higher-orders in $\lambda$, different deformations, and deformations in larger sectors of the theory - they all ultimately require the exact calculation of the form factors for the AdS string world-sheet theory. Such quantities would provide an alternative method for computing planar gauge-theory structure constants, or equivalently the string vertex operator \cite{Bajnok:2015hla} which satisfies a similar set of axioms, and would also provide a means for computing the world-sheet S-matrix for deformed theories to all orders in $\lambda$.  One approach to the computation of form factors is the free field representation  developed by Lukyanov \cite{Lukyanov:1993pn} (see also \cite{Lukyanov:1992sc}) which has been successfully applied to a range of models, for example the SU(2) Thirring and sine-Gordon models \cite{Lukyanov:1993pn}, the O(3) non-linear sigma-model \cite{Horvath:1994rg}, the SU(N) Gross-Neveu models \cite{Britton:2013uta} and, of particular relevance to the string world-sheet theory,  the principal chiral model with a product group structure \cite{Frolov:2017ird}. 

A semi-classical approach to studying deformations of the AdS$_5 \times $S$^5$ geometry, being valid at large $g$, would be complementary to the methods considered here. The classical world-sheet theory in deformed backgrounds, for example the marginal deformations discussed above but also black-hole geometries, will no longer be integrable but in those cases where there is a parameter that can be taken small one may attempt to use techniques, based upon the inverse scattering transform or related methods, previously used for nearly integrable systems \cite{kaup1976perturbation,karpman1977pis, karpman1977perturbation} to construct classical solutions and compute their charges. Given the relation between deformations and FFPT such methods may also be useful for studying world-sheet form factors semi-classically \cite{Bajnok:2014sza, Bajnok:2016xxu}.

\pagebreak 

\section*{Acknowledgements}
We would like to thank Y. Jiang and S. Frolov for useful comments. This work was supported by SFI grant 15/CDA/3472 and Marie Curie Grant CIG-333851.

\appendix

\section{Higher-Order Potential Terms}
\label{app:HOPot}
We record here the quartic and sextic terms of the potential to $\order(g^4)$ which are used to compute the Feynman rules. 
Quartic terms up to order $g^4$:
\< 
\label{eq:quartic_pot}
V_{\rm quartic}&=&   \frac{b_0}{2}(\varphi^\ast{}^2 (\partial_x \varphi)^2+\varphi^2(\partial_x \varphi^\ast)^2)
+\frac{g^2}{2}\Big[ b_1 \left(
\varphi^2 (\partial_x^2\varphi^\ast)^2 \right.\nn\\
&+&\left. \varphi^\ast{}^2 (\partial_x^2\varphi)^2
+8 \partial_x\varphi \partial_x\varphi^\ast{}(\varphi^\ast \partial_x^2 \varphi+\varphi \partial_x^2 \varphi^\ast )\right)
+4(3b_1+2 b_2) (\partial_x\varphi)^2 (\partial_x\varphi^\ast)^2\Big] \nn \\ 
&-& g^4 \Big[ b_3 \left(\varphi^\ast{}^2 (\partial_x^3 \varphi)^2 +(\partial_x^3\varphi^\ast)^2 \varphi^2 +18 (\partial_x\varphi^\ast)^2 (\partial_x^2\varphi)^2 +18 (\partial_x^2\varphi^\ast)^2 (\partial_x\varphi)^2 \right. \nn\\ 
& &\kern+65pt  +6 (\varphi  \partial_x\varphi^\ast-\varphi^\ast  \partial_x\varphi) \left( \partial_x^3\varphi^\ast \partial_x^2\varphi - \partial_x^3\varphi \partial_x^2\varphi^\ast\right) \nn\\ 
& &\kern+65pt 
 +6 \partial_x^3\varphi^\ast \varphi  \partial_x^2\varphi^\ast \partial_x\varphi +6 \varphi^\ast  \partial_x^3\varphi \partial_x\varphi^\ast \partial_x^2\varphi  \nn\\ 
 & &\kern+65pt \left.
+6 \partial_x\varphi \partial_x\varphi^\ast  \left(6 \partial_x^2\varphi^\ast \partial_x^2\varphi- \partial_x^3\varphi^\ast \partial_x\varphi - \partial_x^3\varphi \partial_x\varphi^\ast \right)    \right)\nn\\ 
& &\kern+30pt 
+8 b_4 \,\partial_x\varphi^\ast \partial_x^2\varphi^\ast \partial_x\varphi \partial_x^2\varphi +2 b_5 \left(\partial_x\varphi^\ast \partial_x^2\varphi+\partial_x^2\varphi^\ast  \partial_x\varphi\right)^2\Big] +O(g^6)
\>

Sextic terms up to order $g^4$:
\<
\label{eq:sextic_pot}
V_{\rm sextic}&=&-\frac{b_0}{4}\varphi \varphi^\ast(\varphi^\ast \partial_x \varphi +\varphi \partial_x \varphi^\ast)^2+\frac{g^2}{2}
\Big[ b_1 \left(8 |\varphi|^2|\partial_x \varphi|^4\right. \nn\\
& &\kern+40pt +\varphi^\ast \partial_x^2 \varphi(2|\varphi|^2|\partial_x \varphi|^2+\tfrac{1}{2} |\varphi|^2\varphi^\ast \partial_x^2 \varphi-3\varphi^\ast{}^2(\partial_x\varphi)^2\nn\\
& &\kern+40pt\left. +\varphi \partial_x^2 \varphi^\ast(2|\varphi|^2|\partial_x \varphi|^2+\tfrac{1}{2} |\varphi|^2\varphi \partial_x^2 \varphi^\ast -3\varphi^2(\partial_x\varphi^\ast)^2\right)\nn\\
& &\kern+20pt+8 b_2 (\varphi^\ast{}^2 (\partial_x\varphi)^2+\varphi^2 (\partial_x\varphi^\ast)^2)\Big] \nn\\
&+& \frac{g^4}{2} \Big[-b_3 \left(2 \varphi^\ast{}^2 \partial_x^3\varphi^\ast \partial_x^3\varphi \varphi ^2-3 \partial_x^3\varphi \varphi ^2 (\partial_x\varphi^\ast)^3 +3 \varphi^\ast  \partial_x^3\varphi \right. \nn\\ 
& &\kern+50pt + \varphi^2 \partial_x\varphi^\ast \partial_x^2 \varphi^\ast+\varphi^\ast{}^3 \partial_x^3 \varphi^2 \varphi +36 \varphi^\ast  \varphi  (\partial_x\varphi^\ast)^2 (\partial_x^2 \varphi)^2 \nn\\ 
& &\kern+50pt +9 \varphi^\ast  \partial_x^3 \varphi^\ast \varphi ^2 \partial_x\varphi^\ast \partial_x^2 \varphi +36 \varphi ^2 (\partial_x \varphi^\ast)^2 \partial_x^2 \varphi^\ast \partial_x^2 \varphi-3 \varphi^\ast{}^2 \partial_x^3 \varphi^\ast (\partial_x\varphi)^3  \nn\\ 
& &\kern+50pt +36 \varphi^\ast  \varphi  (\partial_x^2 \varphi^\ast)^2 (\partial_x\varphi)^2 +36 (\partial_x\varphi^\ast)^3 (\partial_x\varphi)^3 +9 \varphi^\ast  \partial_x^3 \varphi^\ast \varphi ^2 \partial_x^2 \varphi^\ast \partial_x\varphi  \nn\\ 
& &\kern+50pt -3 \partial_x^3 \varphi^\ast \varphi ^2 (\partial_x\varphi^\ast)^2 \partial_x\varphi -6 \varphi^\ast  \partial_x^3 \varphi^\ast \varphi \partial_x \varphi^\ast (\partial_x \varphi)^2 +36 \varphi ^2 \partial_x \varphi^\ast (\partial_x^2 \varphi^\ast)^2 \partial_x \varphi  \nn\\ 
& &\kern+50pt +72 \varphi  (\partial_x \varphi^\ast)^2 \partial_x^2 \varphi^\ast (\partial_x \varphi)^2 +36 \varphi^\ast  \partial_x \varphi^\ast \partial_x^2 \varphi^\ast (\partial_x\varphi)^3 +9 \varphi^\ast{}^2 \partial_x^3 \varphi \varphi  \partial_x \varphi^\ast \partial_x^2 \varphi  \nn\\ 
& &\kern+50pt +9 \varphi^\ast{}^2 \partial_x^3 \varphi \varphi  \partial_x^2 \varphi^\ast \partial_x \varphi -6 \varphi^\ast  \partial_x^3 \varphi \varphi  (\partial_x \varphi^\ast)^2 \partial_x \varphi -3 \varphi^\ast{}^2 \partial_x^3 \varphi \partial_x \varphi^\ast (\partial_x \varphi)^2  \nn\\ 
& &\kern+50pt +3 \varphi^\ast{}^2 \partial_x^3 \varphi^\ast \varphi  \partial_x \varphi \partial_x^2 \varphi +36 \varphi^\ast{}^2 \partial_x^2 \varphi^\ast (\partial_x\varphi)^2 \partial_x^2 \varphi +36 \varphi  (\partial_x \varphi^\ast)^3 \partial_x \varphi \partial_x^2 \varphi  \nn\\ 
& &\kern+50pt +36 \varphi^\ast{}^2 \partial_x \varphi^\ast \partial_x \varphi (\partial_x^2 \varphi)^2 +72 \varphi^\ast  (\partial_x \varphi^\ast)^2 (\partial_x \varphi)^2 \partial_x^2 \varphi +\varphi^\ast  (\partial_x^3 \varphi^\ast)^2 \varphi ^3  \nn\\ 
& &\kern+50pt  \left. +108\varphi^\ast  \varphi  \partial_x \varphi^\ast \partial_x^2 \varphi^\ast \partial_x \varphi \partial_x^2 \varphi  +3 \varphi^\ast{}^3 \partial_x^3 \varphi \partial_x \varphi \partial_x^2 \varphi+3 \partial_x^3\varphi^\ast \varphi ^3 \partial_x \varphi^\ast \partial_x^2 \varphi^\ast\right)
\nn\\ 
& &\kern+30pt -b_4 \left(8 \partial_x^2 \varphi^\ast \partial_x^2 \varphi \left(\varphi^\ast{}^2 (\partial_x \varphi)^2+\varphi ^2 (\partial_x \varphi^\ast)^2\right) \right. \nn\\ 
& &\kern+50pt +8 \partial_x \varphi^\ast \partial_x \varphi \left(\varphi ^2 (\partial_x^2 \varphi^\ast)^2  +\varphi^\ast{}^2 (\partial_x^2 \varphi)^2-2 \varphi  (\partial_x \varphi^\ast)^2 \partial_x^2 \varphi-2 \varphi^\ast  \partial_x^2 \varphi^\ast (\partial_x \varphi)^2 \right.\nn\\ 
& & \kern+110pt  \left.\left. \,+4 \varphi  \partial_x \varphi^\ast \partial_x^2\varphi^\ast \partial_x \varphi +4 \varphi^\ast  \partial_x \varphi^\ast \partial_x \varphi \partial_x^2 \varphi +8 (\partial_x \varphi^\ast)^2 (\partial_x \varphi)^2  \right)\right) \nn\\ 
& &\kern+30pt -8 b_5 \left(\partial_x \varphi^\ast \partial_x^2 \varphi+\partial_x^2 \varphi^\ast \partial_x\varphi \right) \left(\varphi (\partial_x \varphi^\ast)^2 \partial_x \varphi+\varphi^\ast  \partial_x \varphi^\ast (\partial_x \varphi)^2 \right. \nn\\ 
& &\kern+180pt \left.+(\varphi^\ast)^2 \partial_x \varphi \partial_x^2 \varphi+\varphi ^2 \partial_x \varphi^\ast \partial_x^2 \varphi^\ast\right) \nn\\ 
& &\kern+30pt-64 b_6 (\partial_x \varphi^\ast)^3 (\partial_x \varphi)^3\Big] +O(g^6)
\>

\bibliographystyle{nb}
\bibliography{FormFactors}

\end{document}